\documentclass[10pt,journal,letterpaper,compsoc]{IEEEtran}

\usepackage{times,epsfig,epstopdf}
\usepackage{graphicx,epsfig,color,endnotes,alltt}
\usepackage{url}
\usepackage{multirow,array}
\usepackage[noend]{algorithmic}
\usepackage{algorithm}
\usepackage{subfigure}
\usepackage{amssymb,amsmath}
\newcommand{\ameliezhou}[1]{{ #1}}

\begin{document}

\pagestyle{plain}
%
\title{Monetary Cost Optimizations for Hosting Workflow-as-a-Service in IaaS Clouds}


%
%
\author{Amelie~Chi~Zhou, Bingsheng~He and Cheng~Liu \\
~{Nanyang Technological University} }



\IEEEcompsoctitleabstractindextext{%
\begin{abstract}

Recently, we have witnessed workflows from science and other
data-intensive applications emerging on Infrastructure-as-a-Service (IaaS)
clouds, and many workflow service providers offering workflow as a service (WaaS).
The major concern of WaaS providers is to minimize the monetary cost
of executing workflows in the IaaS cloud. While there have been previous studies on
this concern, most of them
assume static task execution time and static pricing scheme, and have the
QoS notion of satisfying a deterministic deadline. However, cloud
environment is \emph{dynamic}, with performance dynamics caused by
the interference from concurrent executions and price dynamics like
spot prices offered by Amazon EC2. Therefore, we argue that WaaS
providers should have the notion
of offering probabilistic performance guarantees for individual
workflows on IaaS clouds. We
develop a probabilistic scheduling framework called
\emph{Dyna} to minimize the monetary cost while offering
probabilistic deadline guarantees. The framework includes an
$\mathit{A^\star}$-based instance configuration method for performance dynamics,
and a hybrid instance
configuration refinement for utilizing
spot instances. Experimental
results with three real-world scientific workflow applications on
Amazon EC2 demonstrate (1) the accuracy of our
framework on satisfying the probabilistic deadline guarantees required
by the users; (2) the
effectiveness of our framework on reducing monetary cost in comparison with the
existing approaches.

\end{abstract}


\begin{keywords}
Cloud computing, cloud dynamics, spot prices, monetary cost optimizations, scientific workflows.
\end{keywords}}

\maketitle

\IEEEdisplaynotcompsoctitleabstractindextext

%
\IEEEpeerreviewmaketitle

\section{Introduction}\label{sec:intro}

\IEEEPARstart{C}{loud} computing has become a popular computing
infrastructure for various applications. One attractive feature of
cloud computing is the pay-as-you-go charging scheme, where users
only need to pay for the actual consumption of storage and
computation hours. This feature unlocks the opportunities of
large-scale computation without physically owning a cloud. Recently,
we have witnessed many workflows from various
scientific and data-intensive applications
deployed and hosted on the Infrastructure-as-a-Service (IaaS) clouds
such as Amazon EC2 and other cloud
providers~\cite{amazonCaseStudies,msCaseStudies}. In those
applications, workflows are submitted and
executed in the cloud and each workflow is usually associated with a
deadline for QoS purposes~\cite{SC11,SIGMOD11}.
This has formed a new software-as-a-service model for hosting
workflows in the cloud, and we refer it as Workflow-as-a-Service (WaaS).
WaaS providers charge users based on the execution of their workflows
and QoS requirements. On the other hand, WaaS providers rent cloud
resources from IaaS clouds, which induces the monetary cost.
Monetary cost and performance
are usually the two most important optimization factors
for hosting WaaS for the providers in the cloud.
In this paper, we investigate whether and how WaaS providers
can reduce the monetary cost of hosting WaaS while offering
performance guarantees for individual workflows.

Monetary cost optimizations have been classic research topics in grid
and cloud computing environments.
Over the era of grid computing,
cost-aware optimization techniques have been extensively studied.
Researchers have addressed various problems: minimizing cost given
the performance requirements~\cite{Yu:2005:CSS:1107836.1107867},
maximizing the performance for given budgets~\cite{Sakellariou07}
and scheduling optimizations with both cost and performance
constraints~\cite{Duan2007}. When it comes to cloud computing, the
pay-as-you-go pricing, virtualization and elasticity features of
cloud computing open up various challenges and
opportunities~\cite{SC11,scheduling1}. For example,
most cloud providers offer instance hour billing model. Partial-hour
consumption is always rounded up to one hour. Although some other billing
models have been proposed (e.g., Google's IaaS service charges by minutes of use),
hourly billing is still the most commonly adopted model.
Recently, there have
been many studies on monetary cost optimizations with resource
allocations and task scheduling according to the features of cloud
computing (e.g.,
~\cite{SC11,SIGMOD11,scheduling1,scheduling2,infocom12,cloudcom11,Malawski:2012:CDP:2388996.2389026}).
Although the above studies have demonstrated their effectiveness in
reducing the monetary cost, all of them assume static task execution
time and consider only fixed pricing scheme (only \emph{on-demand}
instances in Amazon's terminology). Particularly, they have the
following limitations.

First, cloud is by design a shared infrastructure. The resources in
the cloud, such as the computation, storage and network resources,
are shared by many concurrent jobs/tasks. Previous
studies~\cite{Runtimemeasure,Iosup:2011:PAC:1990761.1990853} have
demonstrated significant variances on I/O and network performance.
The assumption of static task execution time in the previous studies
(e.g.,
~\cite{SC11,SIGMOD11,scheduling1,scheduling2,infocom12,cloudcom11,Malawski:2012:CDP:2388996.2389026})
does not hold in the cloud. 
\ameliezhou{Under the static execution time assumption, 
existing cost optimizations and algorithms try to satisfy the 
conventional QoS notion of a soft deadline with 100\% guarantee.
We denote the conventional deadline notion as the ``deterministic deadline'' 
in the remainder of this paper.
However, due to performance dynamics, the actual execution time of a job 
after optimizations can be varying values with different probabilities. 
Requiring all the varying execution time to meet the deterministic 
deadline is costly and meaningless.
Thus, the deterministic deadline notion is not desirable to offer 
performance guarantees in dynamic cloud environments and the existing 
studies need to be revisited and adapted to performance dynamics.}

Second, cloud, which has evolved into an economic
market~\cite{HOTCLOUD10}, has dynamic pricing. Amazon EC2 offers
spot instances, whose prices are determined by market demand and
supply. Previous
studies~\cite{SC11,Yu:2005:CSS:1107836.1107867,Garg:2010:TCT:1838759.1838827,
Sakellariou07schedulingworkflows,SIGMOD11,BagofTasks} consider fixed
pricing schemes only and their results need revisits in the
existence of spot instances. On the other hand, spot instances can
be used to reduce monetary cost~\cite{INRIA,
DesicionModel,spotHotCloud10,hybridPeakload,SIPerformanceAvailability,DBLP:journals/corr/abs-1110-5969},
because the spot price is usually much lower than the price of
on-demand instances of the same type. However, a spot instance may
be terminated at any time when the bidding price is lower than the
spot price (i.e., out-of-bid events). The usage of spot instances
may cause excessive long latency due to failures. Most of the
previous studies do not consider deadline constraints of individual
workflows.

Those two kinds of cloud dynamics make challenging the problem of
minimizing the cost of WaaS providers while satisfying QoS requirements of
individual workflows. They add two new dimensions to the problem,
and dramatically increase the solution space (see
Section~\ref{sec:alg}). Moreover, the deterministic deadline notion 
will lead the optimizations to worst-case
performance/price prediction. Even worse, worst-case predictions can
be unknown or unpredictable in some cases (e.g., there are some
exceptionally high spot prices in the price history of Amazon EC2).

In order to address performance and price dynamics, we define the
notion of probabilistic performance guarantees to represent QoS.
Each workflow is associated with a probabilistic deadline guarantee of $p$\%.
Deterministic deadline guarantee can be viewed as a special case of probabilistic
deadline guarantee of 100\%.
WaaS provider guarantees that the workflow's execution time is
at the $p$-th percentile of the distribution of the workflow
execution time in the dynamic cloud environment.
This is just like many IaaS cloud providers offer a resource
availability guarantee of 99.95\%~\cite{ec2availability}.
Under this notion, we propose a probabilistic framework called \emph{Dyna} to
minimize the cost of the WaaS provider while satisfying the probabilistic
performance guarantees of individual workflows predefined by the user. The
framework embraces a series of static and dynamic optimizations for
monetary cost optimizations, which are specifically designed for
cloud dynamics. We develop probabilistic models to capture the
dynamics in I/O and network performance, and spot
prices~\cite{Deconstructing,ModelSpot}.
We further propose a hybrid
execution with both spot and on-demand instances, where spot
instances are adopted to potentially reduce monetary cost and
on-demand instances are used as the last defense to meet deadline
constraints.

We calibrate the cloud dynamics from a real cloud
provider (Amazon EC2) for the probabilistic models on I/O and
network performance as well as spot prices.
We perform experiments using three real-world workflow applications
\ameliezhou{on real cloud environment}.
Our experimental results demonstrate the following two major results.
\begin{enumerate}
  \item With the calibrations from
Amazon EC2, Dyna can accurately capture the cloud dynamics
and obtain optimization results always with the same or slightly
better probability than the probabilistic
performance guarantees required by the users.
  \item The hybrid
instance configuration approach significantly reduces the monetary
cost by 15--73\% over other state-of-the-art algorithms which only adopt
on-demand instances.
\end{enumerate}
\noindent To the best of our knowledge, our study is the first
probabilistic scheduling framework for WaaS providers in the
cloud.

The rest of the paper is organized as follows.
We formulate our problem and review the
related work in Section~\ref{sec:problem}. We present our detailed
framework design in Section~\ref{sec:alg}, followed by the
experimental results in Section~\ref{sec:eval}. Finally, we conclude
this paper in Section~\ref{sec:conclusion}.

\section{Background and Related Work}\label{sec:problem}
In this section, we present the application scenario and describe
the terminology, followed by the related work.

\subsection{Application Scenario}

\begin{figure}
  \centering
  \includegraphics[width=0.5\textwidth]{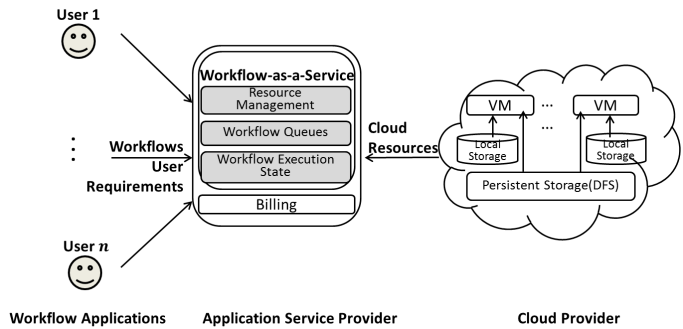}
  \caption{Application scenario of this study.}\label{fig:water}
\end{figure}

Figure~\ref{fig:water}
illustrates our application scenario.
In this study, we consider a typical scenario of offering software-as-a-service
model for workflows on IaaS clouds~\cite{SC11,amazonCaseStudies}.
We call this model Workflow-as-a-Service (WaaS).
In this hosting, different application owners submit a number of
workflow classes with different parameters.
A workflow class represents the template of the same workflow structure
with different input data parameters. WaaS providers allow users to instantiate a workflow execution by specifying the input data to a workflow class,
with specified deadlines for QoS purposes.
The WaaS providers charge users according to the workflow
classes (which reflect the complexity of the workflow) and QoS requirements.
We can see many applications with different workflow structures and purposes in the
cloud~\cite{amazonCaseStudies,msCaseStudies}.
Users (e.g., scientists and officials) can submit their
simulation tasks for predictions, or perform sensitivity analysis.
Users can also perform data analysis on scientific data with data
mining or machine learning techniques.

Different workflow scheduling and resource provisioning algorithms
can result in significant differences in the monetary cost
of WaaS providers running the service on IaaS clouds.
Considering the cloud dynamics, our goal is to provide a probabilistic
scheduling framework to WaaS providers, aiming at minimizing the
monetary cost while satisfying users' QoS requirements.

\subsection{Terminology}\label{subsec:term}

{\bf Instance.} An instance is a virtual machine offered by the
cloud provider. Instance in the same type have the same
amount of resources such as CPUs and RAM and the same capabilities
such as CPU speed, I/O speed and network bandwidth.

The instance acquisition has a non-ignorable acquisition time. For
simplicity, we assume the acquisition time is a constant,
$\mathit{lag}$.

An instance can be on-demand or spot. We adopt the instance
definition of Amazon EC2. Amazon adopts the hourly billing
model, where any partial hour of instance usage is rounded to one hour.
Both on-demand and spot instances can be terminated when users no
longer need them. If an instance is terminated by the user, the user has to
pay for any partial hour (rounded up to one hour). For a spot
instance, if it is terminated due to an out-of-bid event, users do not
need to pay for any partial hour of usage.

Table~\ref{tb:spot} shows some statistics of the price history of
four types of spot instances on Amazon in the US East region
during August 2013. We also show the price of the on-demand
instances for these four types. We have the following observations:
a) The spot instances are usually cheaper than on-demand instances.
There are some ``outlier" points where the maximum price is much
higher than the on-demand price. b)
Different types have different variations on the price. These
observations are consistent with the previous
studies~\cite{ModelSpot,Deconstructing}.

\begin{table}
  \centering\begin{small}
  \caption{Statistics on spot prices (\$/hour, August 2013, US East Region) and on-demand prices of Amazon EC2. }\label{tb:spot}
  \begin{tabular}{|c|c|c|c|c|c|c|c|}
    \hline
    Instance type&Average&$stdev$&Min&Max&OnDemand\\\hline
    m1.small&0.048&0.438&0.007&10&0.06\\\hline
    m1.medium&0.246&1.31&0.0001&10&0.12\\\hline
    m1.large&0.069&0.770&0.026&40&0.24\\\hline
    m1.xlarge&0.413&2.22&0.052&20&0.48\\\hline
\end{tabular}
\end{small}
\end{table}

{\bf Job.} A job is expressed as a workflow of tasks with precedence constraints.
A job has a deadline, which is a ``soft deadline'', unlike
a ``hard deadline" as in hard real-time systems.
In this study, we consider the deadline of a job as a probabilistic requirement.
Suppose a workflow is specified with a probabilistic deadline guarantee of $p$\%.
WaaS provider guarantees that the workflow's execution time is at the $p$-th
percentile of the distribution of the workflow execution time in the
dynamic cloud environment.

{\bf Task.} A task can be of different classes, e.g., compute-intensive
and I/O-intensive tasks, according to the dominating part of the total
execution time. The execution time (or response time) of a task
is usually estimated using estimation methods such as task profiling, machine
benchmarking and statistical analysis.
\ameliezhou{In this study, we use a simple performance estimation model on predicting
the task execution time. Since workflows are often regular and predictable~\cite{SC11,Yu:2005:CSS:1107836.1107867},
this simple approach is already sufficiently accurate in practice,
as shown in our experiments on real cloud environments.
Specifically,} we estimate the
execution time of a task on different types of instances using the following
profile: $<$$\#\mathit{instr}$,
$\mathit{d_{seqIO}}$, $\mathit{d_{rndIO}}$, $\mathit{netInput}$,
$\mathit{netOutput}$$>$, where $\#\mathit{instr}$ represents the
total number of instructions to be executed for the task,
$\mathit{d_{seqIO}}$ and $\mathit{d_{rndIO}}$ are the amount of I/O
data for sequential and random accesses respectively to local disk,
and $\mathit{netInput}$ and $\mathit{netOutput}$ are the amount of
input and output data that need to be read in and sent out
respectively. We estimate the task execution time as the summation of
its CPU time, I/O time and networking time.

In the estimation, the CPU time is determined by the $\#\mathit{instr}$
of the task as well as the CPU frequency of the instance that the task
is executed on. Similarly, the I/O time and networking time
is determined by the I/O and networking data size
divided by the I/O and network bandwidth, respectively.
We model the I/O and network performance in the cloud as probabilistic
distributions and use them to estimate the dynamic execution time of tasks
in the cloud.

{\bf Instance configuration.} The {\em hybrid instance configuration} of
a task is defined as an $n$-dimension vector: $<$
($\mathit{type_1}$, $\mathit{price_1}$, $\mathit{isSpot_1}$),
($\mathit{type_2}$, $\mathit{price_2}$, $\mathit{isSpot_2}$), ...,
($\mathit{type_n}$, $\mathit{price_n}$, $\mathit{isSpot_n}$) $>$,
meaning the task is potentially to be executed by $n$ instances belonging
to $\mathit{type_i}$ ($1\le i \le n$) in sequential. $\mathit{isSpot_i}$
indicates whether the instance is spot or on-demand. If the instance
$i$ is a spot instance, $\mathit{price_i}$ is the specified bidding
price, and the on-demand price otherwise. In our hybrid instance
configuration, only the last dimension of the configuration is
on-demand instance and all previous dimensions are spot instances.
This is because the on-demand instance guarantees 100\% of success execution.

In the hybrid execution of spot and
on-demand instances, a task is initially assigned to a spot instance
of the type indicated by the first dimension of its configuration (if any).
If the task fails on this spot instance, it will be re-assigned to an
instance of the next type indicated by its configuration until it successfully finishes.
Since the last dimension is an on-demand instance type,
the task can always finish the execution,
even when the task fails on all previous spot instances.

\subsection{Related Work}\label{sec:related}

There are a lot of works related to
our study, and we focus on the most relevant ones on cost
optimizations and cloud performance dynamics. To the best of our
knowledge, this work is the first scheduling framework that
captures both performance and price dynamics in the cloud.

{\bf Cost-aware optimizations.}
The pay-as-you-go nature of cloud computing attracts many research
efforts in dynamic resource provisioning. Dynamic virtual machine
provisioning has been determined by control
theory~\cite{Zhu:2010:RPB:1851476.1851516,cloudnet12}, machine
learning~\cite{ZhangICAC} and
models~\cite{Sharma:2011:CEP:2014697.2014786}. On the other hand,
workflow scheduling with deadline and budget constraints
(e.g.,~\cite{Yu:2005:CSS:1107836.1107867,Garg:2010:TCT:1838759.1838827,
Sakellariou07schedulingworkflows,SIGMOD11,BagofTasks,SC12:pegasus,10.1109/TPDS.2013.238})
has been widely studied. Yu et
al.~\cite{Yu:2005:CSS:1107836.1107867} proposed deadline assignment
for the tasks within a job and used genetic algorithms to find
optimal scheduling plans. Kllapi et al. ~\cite{SIGMOD11} studied the
tradeoff between monetary cost and performance, and modeled the
tradeoff as {\em sky line} operations in databases. Those studies
only consider a single workflow with on-demand instances only.
Malawski et al.~\cite{SC12:pegasus} proposed dynamic
scheduling strategies for workflow ensembles. The previous
studies~\cite{SC11,AutoScaleModel,ProfileScale} proposed
auto-scaling techniques based on static execution
time of individual tasks. In comparison with the previous works, the
unique feature of Dyna is that it targets at offering probabilistic
performance guarantees as QoS, instead of deterministic deadlines. Dyna
schedules the workflow by explicitly capturing the performance
dynamics (particularly for I/O and network performance) in the
cloud. Buyya et al.~\cite{10.1109/TPDS.2013.238} proposed an algorithm
with task replications to increase the likelihood of meeting deadlines.

Due to their ability on reducing monetary cost,
Amazon EC2 spot instances have recently
received a lot of interests. Related work can be roughly divided
into two categories: modeling spot prices~\cite{Deconstructing,ModelSpot} and leveraging spot
instances~\cite{INRIA,
DesicionModel,spotHotCloud10,hybridPeakload,SIPerformanceAvailability,DBLP:journals/corr/abs-1110-5969,europar2012/Simon}.

For modeling spot prices, Yehuda et al.~\cite{Deconstructing}
conducted reverse engineering on the spot price and figured out a
model consistent with existing price traces. Javadi et
al.~\cite{ModelSpot,Javadi:2013:CSP:2435448.2435538} developed statistical models for different spot
instance types. Those models can be adopted to our hybrid execution.

For leveraging spot instances, Chohan et al.~\cite{spotHotCloud10}
proposed a method to utilize the spot instances to speed up the
MapReduce tasks, and suggested that fault tolerant mechanisms are
essential to run MapReduce jobs on spot instances. Yi et
al.~\cite{INRIA} introduced some checkpointing mechanisms for
reducing cost of spot instances. Ostermann et al.~\cite{europar2012/Simon}
utilize spot instances for large workflow executions when the Grid resources
are not sufficient. Further
studies~\cite{SIPerformanceAvailability,DBLP:journals/corr/abs-1110-5969}
used spot instances with different bidding strategies and
incorporating with fault tolerance techniques such as checkpointing,
task duplication and migration. Those studies are with spot instance
only, without offering any guarantee on meeting the workflow
deadline like Dyna.

{\bf Cloud performance dynamics.}
A few studies have evaluated the performance of cloud services from
different
aspects~\cite{suhailrehman:initial,Runtimemeasure,Iosup:2011:PAC:1990761.1990853}.
Our calibration results on Amazon EC2 is consistent with Schad et
al.'s work~\cite{Runtimemeasure}. Iosup et
al.~\cite{Iosup:2011:PAC:1990761.1990853} did a cross-platform
comparison on four commercial cloud providers (Amazon EC2, GoGrid,
ElasticHosts and Mosso) for scientific computing workloads. There
have been more in-depth performance studies on specific components (network
performance~\cite{DBLP:journals/corr/abs-1205-1622,Wang:2010:IVN:1833515.1833691}
and I/O
interference~\cite{conf/IEEEcloud/PuLMSKP10,10.1109/TSC.2012.2}).

There have been some proposals to reduce the performance
interference and unpredictability in the cloud, such as network
performance~\cite{Ballani:2011:TPD:2018436.2018465} and I/O
performance~\cite{hovestadt.2011.datacloud,Lin+:HotCloud12,Chiang:2011:TIS:2063384.2063447}.
However, by the design of cloud computing, cloud is shared by many
concurrent executions. Therefore, the performance dynamics caused by
the resource interference is unavoidable.
This paper offers a probabilistic notion to capture the performance
and cost dynamics, and further develop a probabilistic scheduling
framework to minimize the monetary cost with the consideration of those dynamics.

%
%

\section{Framework Design and Implementation}\label{sec:alg}

We first present an overview of the Dyna
framework and then discuss the design details about the optimization
techniques adopted in Dyna.

\subsection{Framework Overview}
When we design the framework for minimizing the monetary cost of WaaS provider, we consider the following design principles.
\begin{itemize}
 \item {\bf Effectiveness.} The proposed framework should optimize the monetary cost with the special consideration on the performance and price dynamics in the cloud environment. Moreover, while satisfying probabilistic deadline guarantees for individual workflows, the framework should be able to find the solution that is comparable or close to the optimal solution.
 \item {\bf Generality.} The reason that we develop a general
    framework has two folds. First, there have been some classic
    job/task scheduling optimizations for monetary cost and
    performance~\cite{Yu:2005:CSS:1107836.1107867,Sakellariou07,Duan2007}.
    A framework allows integrating existing optimizations in a holistic
    manner. Second, different cloud providers may have different
    behaviors and offerings in performance and price dynamics. A general
    framework allows more flexibility in implementing specific
    algorithms for different cloud providers, without affecting other
    key optimization components.
\item {\bf Low runtime overhead.} The cost optimization is an online process and should be lightweight. We should find a good balance between the quality of monetary cost optimization and the runtime overhead of the optimization process itself. Due to the huge space, a thorough exploration of the optimization space is impractical.
\end{itemize}

\begin{figure}
\centering{
  \includegraphics[width=0.4\textwidth]{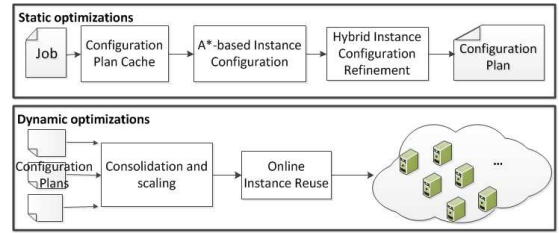}}
 \caption{Static and dynamic optimizations in the Dyna framework}
\label{BigPicture}       
\end{figure}

With the three principles, we propose Dyna, a probabilistic
scheduling framework for workflows. The main components of Dyna are illustrated in
Figure~\ref{BigPicture}. The framework consists of static
and dynamic optimizations.

When a workflow is
submitted to the WaaS provider with the pre-defined probabilistic deadline
guarantee, we first determine the most cost-efficient
configuration plan for each task as static optimizations.
There are two major static optimizations for generating the configuration plan
for a workflow: firstly an $A^\star$-based instance configuration approach
for selecting the on-demand instance type, and secondly the hybrid instance
configuration refinement for considering spot instances.
Note that for the same workflow class (with the same structure and
task profiles), we only need to do the static optimization once,
and we can retrieve the configuration plan directly from the configuration plan cache.
The function of configuration plan cache is to store the configuration plans
for the workflow classes that have been executed.
At runtime, the tasks of the workflow are scheduled according to
their instance configuration and several dynamic optimization techniques
including consolidation and instance reuse are applied to further reduce cost.

In the remainder of this section, we outline the design of static
and dynamic optimizations, and discuss on the implementation details.

\subsection{Static Optimizations}
\label{subsec:static}
The overall functionality of static optimizations is to determine
the suitable instance configuration for each task in the workflow
so that the monetary cost is minimized while the probabilistic
performance guarantees are satisfied.
Ideally, one can consider on-demand and spot instances together.
However, this will induce too large solution space.
Therefore, we consider a divide-and-conquer approach on the static optimizations.
The static optimizations include two kinds of optimizations: $A^\star$-based instance
configuration and hybrid instance configuration refinement.
The rationale is to first determine the on-demand instance type
for each task, and then to perform refinement by introducing
hybrid execution configurations to workflow execution so that the monetary cost is further reduced.
Particularly, we formulate the process of determining the on-demand instance
type into an $A^\star$-based approach. Based on the output from the $A^\star$-based
configuration, we determine the hybrid instance configuration as a refinement for each task.

\subsubsection{$A^\star$-based Instance Configuration}
In this optimization, we determine an on-demand instance type for each task in the workflow.
We formulate the process into an $A^\star$-based search problem.
The reason that we choose $A^\star$ search is to take advantage of its pruning capability
to reduce the large search space while targeting at a high quality solution.

\begin{figure}
\centering
  \includegraphics[width=0.48\textwidth]{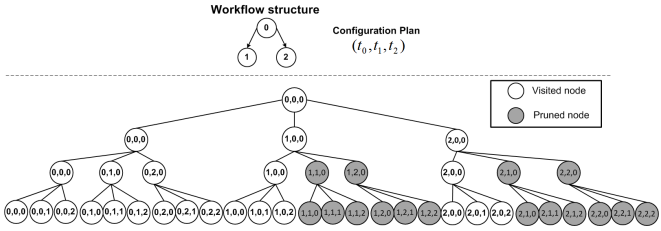}
  \centering
 \caption{An example of the configuration plan search tree in our $A^\star$ algorithm.}
\label{fig:searchtree}       
\end{figure}

The $A^\star$-based instance configuration is extended from the classical $A^\star$ search process.
In the formulated $A^\star$ search, we define the \emph{state} to be a configuration plan
to the workflow.
For ease of presentation, we represent a configuration
plan to be a multi-dimensional vector of the instance configuration for each task in the workflow.
The $i$-th dimension of the vector corresponds to
the assigned instance configuration of the task of ID $i$.
We have two issues to clarify. First, we ensure that each task has a unique ID.
The order of assigning ID to the task does not affect the correctness
of our algorithm. In this study, we simply use a topological order
to assign the task ID. Second, at this static optimization,
the instance configuration consists of a single on-demand instance,
which will be extended to hybrid instance configuration in the next subsection.
For example, as shown in
Figure~\ref{fig:searchtree}, a state is represented
as $(t_0, t_1, t_2)$, meaning that task $i$ ($0\le i \le 2$) is configured
with on-demand instance type $t_i$.

Originally, $A^\star$ search algorithm is a heuristic searching method that searches for the
shortest path from a given initial state to the specified goal state.
During the search process, $A^\star$ algorithm attempts to calculate the
smallest cost so far and to prune the unnecessary states at each state.
Particularly, $A^\star$ evaluates a state $s$ by combining
two distance metrics $g(s)$ and $h(s)$, which are the actual
distance from the initial state to the state $s$ and the estimated distance from
the state $s$ to the goal state, respectively.
$g(s)$ and $h(s)$ are also referred as $g$ score and $h$ score for $s$, respectively.
We estimate the total search cost for $s$ to be $f(s)=g(s)+h(s)$.
The $A^\star$ search algorithm is equivalent to the search and pruning process on a search tree.
Figure~\ref{fig:searchtree} shows an example of the configuration plan search tree
in our problem setting. Each node represents a state in the $A^\star$ search algorithm
(a state is a
configuration plan for the workflow in our paper).
A child state only differs with its parent state,
by replacing a dimension with a more expensive instance type.

\begin{algorithm}
\begin{small}
\caption{$A^\star$-based instance configuration search from initial state $S$ to goal state $D$} \label{alg:a*}
\begin{algorithmic}[1]

\REQUIRE {
$\mathit{Max\_iter}$: Maximum number of iterations; \\
$\mathit{deadline, p_r}$: Required probabilistic deadline guarantee}

\STATE ClosedList := empty;
\STATE OpenList := ${S}$;
\STATE $\mathit{upperBound}$ := 0;

\STATE $\mathit{g[S] := 0}$;
\STATE $\mathit{f[S] := g[S]}$ + $\mathit{estimate\_h\_score(S)}$;
\WHILE {\NOT(OpenList is empty \OR reach $Max\_iter$)}
    \STATE $\mathit{current}$ := the state in OpenList having the lowest $\mathit{f}$ value;
    \STATE $\mathit{percentile}$ := $\mathit{estimate\_performance(current, p_r)}$;
    \IF {$\mathit{percentile <= deadline}$}
        \STATE $\mathit{current\_cost}$ := $\mathit{estimate\_cost(current)}$;
        \IF {$\mathit{current\_cost < upperBound}$}
            \STATE $\mathit{upperBound = current\_cost}$;
            \STATE $D$ = $\mathit{current}$;
        \ENDIF
    \ENDIF
    \STATE Remove $\mathit{current}$ from OpenList;
    \STATE Add $\mathit{current}$ to ClosedList;
    \FOR{\textbf{each} $\mathit{neighbor}$ in neighboring states of $\mathit{current}$ }
        \STATE $\mathit{g[neighbor]}$ := $\mathit{cal\_g\_score(neighbor,S)}$;
        \STATE $\mathit{f[neighbor] := g[neighbor]}$ + $\mathit{estimate\_h\_score(neighbor)}$;
        \IF {$\mathit{f[neighbor] >= upperBound}$ \OR $\mathit{neighbor}$ is in ClosedList}
            \STATE continue;
        \ENDIF
        \IF {$\mathit{neighbor}$ is not in OpenList}
            \STATE Add $\mathit{neighbor}$ to OpenList;
        \ENDIF
    \ENDFOR
\ENDWHILE
\STATE Return $D$;
\end{algorithmic}
\end{small}
\end{algorithm}

The goal of our $A^\star$-based algorithm is to search for an optimal state on the search tree, which has the
lowest monetary cost and can satisfy the probabilistic deadline guarantee.
Algorithm~\ref{alg:a*} shows the optimization process of the $A^\star$-based instance
configuration algorithm.
The algorithm maintains all the states in the search tree with two list structures: ClosedList and OpenList.
The OpenList is to store the states that are being considered
to find the goal state and ClosedList is to store the states that do not
need to consider again during the $A^\star$ search.
Functions $\mathit{estimate\_h\_score}$ and $\mathit{estimate\_g\_score}$ return
the $h$ and $g$ scores of states, respectively, and $\mathit{estimate\_cost}$ and $\mathit{estimate\_performance}$
return the monetary cost and feasibility estimations of states, respectively.
The core operations for the above four functions are to estimate the
probabilistic distributions for the execution time and the monetary
cost of a state (i.e., a configuration plan).

We develop probabilistic
distribution models to describe the performance dynamics for I/O and
network. Previous studies~\cite{Runtimemeasure,HOTCLOUD10} show that
I/O and network are the major sources of performance dynamics in the
cloud due to resource sharing while the CPU performance is rather
stable for a given instance type. We define the probability of the
I/O and network bandwidth equaling to a certain value $x$ on
instance type $\mathit{type}$ to be:
$\mathit{P_{seqBand,type}(seqBand = x)}$,
$\mathit{P_{rndBand,type}(rndBand = x)}$, $\mathit{P_{inBand,
type}(inBand = x)}$ and $\mathit{P_{outBand, type}(outBand = x)}$ as
the probabilistic distributions for the sequential I/O, random I/O,
downloading and uploading network performance from/to persistent storage of instance
type $\mathit{type}$, respectively. In our calibrations on Amazon
EC2, $\mathit{P_{rndBand,type}(rndBand = x)}$ conforms to normal
distributions and the other three conform to Gamma distributions
(Section~\ref{sec:eval}).
Given the I/O and network performance distributions, we manage to model
the execution time of a task on different instance types
with probabilistic distribution functions (PDFs), given the I/O and networking data size.

Having modeled the execution time of tasks as probabilistic distributions,
we now introduce how to estimate the execution cost of tasks.
Consider a task with on-demand instance type
$\mathit{type}$ with the price $\mathit{p}$.
We estimate the expected monetary cost of the task
to be $p$ multiplied by the expected execution time
on the $\mathit{type}$-type on-demand instance.
The expected monetary cost of the entire workflow is the total
cost of all the tasks in the workflow.
Here, we have ignored the rounding monetary cost in the
estimation. This is because in the WaaS environment,
this rounding monetary cost is usually amortized among many tasks. Enforcing
the instance hour billing model could severely limit the
optimization space, leading to a suboptimal solution.

\begin{figure}
\centering
  \includegraphics[width=0.45\textwidth]{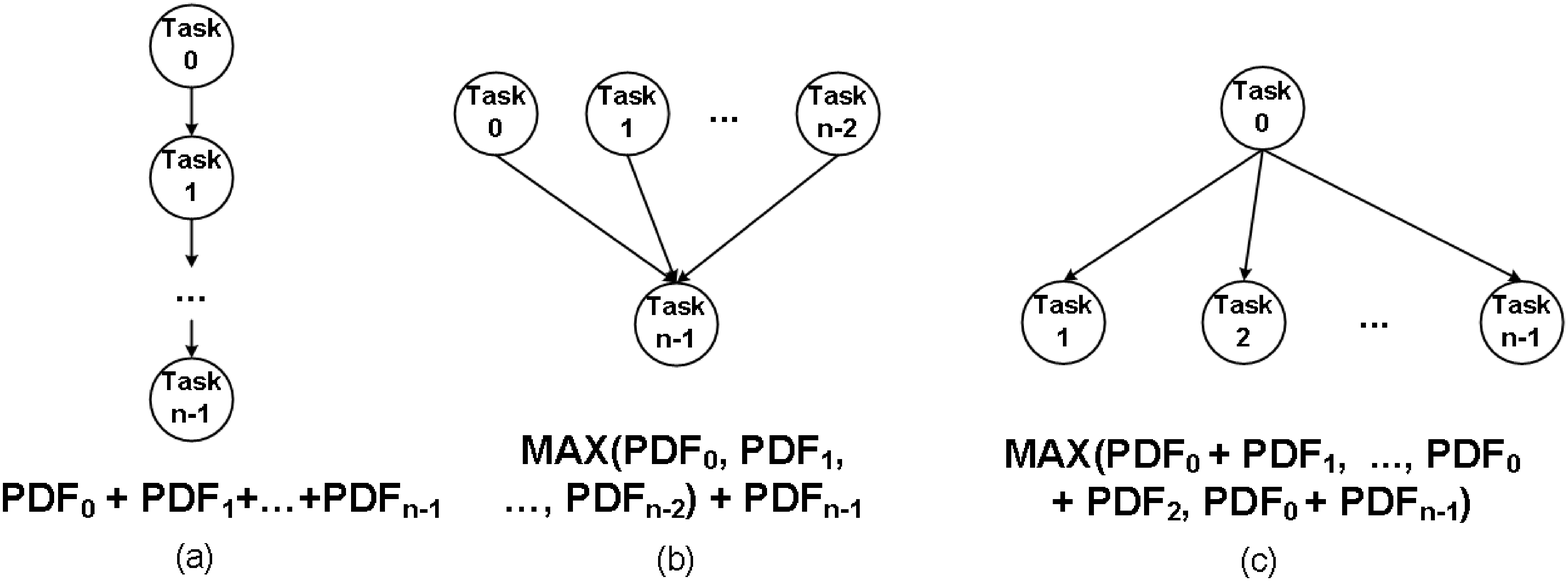}
\centering
  \caption{Basic workflow structures and their probabilistic distributions of the execution time,
  denoting the execution time distribution of Task 0, 1,..., $n-1$ to be PDF$_0$, PDF$_1$, ..., PDF$_{n-1}$, respectively.}
\label{fig:conv}       
\end{figure}

Another core operation is to determine whether a configuration plan
satisfies the probabilistic deadline requirement.
Given the execution time distribution of each task under the evaluated
configuration plan, we first calculate the
execution time distribution of the overall workflow.
Particularly, we decompose the workflow structure into the three basic structures,
as shown in Figure~\ref{fig:conv}.
Each basic structure has $n$ tasks ($n\ge 2$).
The execution time distribution of each basic structure is calculated
with the execution time distributions of individual tasks. For example,
the execution time distribution of the structure
in Figure~\ref{fig:conv}(b) is calculated as $\mathit{MAX(PDF_0,PDF_1,...,PDF_{n-2})+PDF_{n-1}}$,
where $\mathit{PDF_i}$ ($0\le i\le n-1$) is the probabilistic
distribution of the execution time of task $i$.
The ``$+$'' operation of two probabilistic distributions calculates the convolution
of the two distributions
and the $\mathit{MAX}$ operation finds the distribution of the maximum of two random
variables modeled by the two distributions.
After obtaining the execution time distribution of the workflow, we check its
percentile at the required probabilistic deadline guarantee. Only if the returned
percentile is smaller than deadline, the evaluated configuration plan is feasible.

We set the initial state of the $A^\star$ search to configuration $(0,0,...,0)$,
where each task is configured with the cheapest instance type
(instance type $0$). The goal state is the optimal
configuration plan. We define $g(s)$ to be the monetary cost difference
between $s$ and the initial state.
We estimate $h(s)$ to be the monetary cost of configuration plan $s$.
We maintain the lowest cost found during the searching process as the \emph{upper bound} to
prune the unuseful states on the search tree.
Note, we only consider the feasible states that satisfy the probabilistic deadline guarantee.
The leaves of the tree include all possible configuration plans of a workflow.
Starting from the initial configuration, the search tree is constructed by
expanding the state with its child states.
At level $l$, the expanding operation is to replace the $l$-th dimension
in the state with all possible more expensive instance types.
Thus, all the child states have better execution time distribution than its ancestor states.
If the monetary cost
of a state $s$ is higher than the upper bound, its successors are unlikely
to be the goal state since they have more expensive configurations than $s$.
For example, suppose configuration $(1,1,0)$ on the search tree in Figure~\ref{fig:searchtree}
has a higher monetary cost than the upper bound.
The grey nodes on the search tree can be pruned
during the $A^\star$ search.

\subsubsection{ Hybrid Instance Configuration Refinement}
\begin{figure}
\centering
  \includegraphics[width=0.3\textwidth]{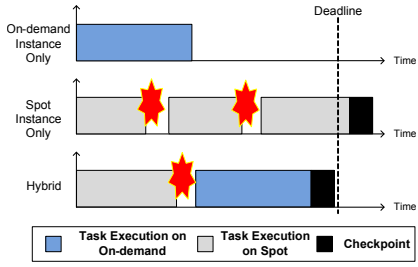}
 \caption{Hybrid Approach Demonstration}
\label{sec3:fig:1}       
\end{figure}

We consider the adoption of spot instances as a refinement to
the configuration plan obtained from the $A^\star$-based instance
configuration algorithm to further reduce monetary cost.
The major issue with adopting spot instances is that,
running a task on spot instances may suffer from the out-of-bid events and fail to
meet the deadline requirements.
We propose a hybrid instance configuration with the adoption of both on-demand
and spot instances to tackle this issue. The basic idea is, if the deadline allows,
we can run a task on a spot instance in advance
(before we run the task on an on-demand instance). If the task
can finish on the spot instance, the monetary cost tends to be lower.
It is possible
that we can try more than one spot instances, if the previous
spot instance fails (as long as it can reduce the monetary cost
and satisfy the probabilistic performance guarantee).
If all spot instances in the hybrid instance configuration fail,
the task is executed on an on-demand instance
to ensure the deadline. Figure~\ref{sec3:fig:1} compares on-demand, spot and hybrid
configurations for a single task. If a task fails on a spot
instance, its failure does not trigger the re-execution of its
precedent tasks. The results of precedent tasks are already checkpointed, and
materialized to the persistent storage in the cloud (such as Amazon S3).
Dyna performs checkpointing only when the task ends, which is simple and has much less overhead than the general checkpointing algorithms~\cite{INRIA}.

\begin{figure}
\centering
  \includegraphics[width=0.3\textwidth]{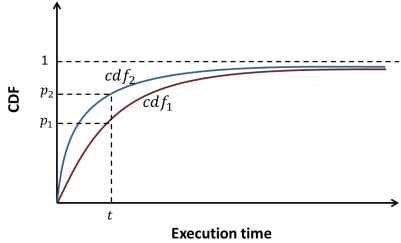}
\centering
  \caption{The definition of configuration $C_2\ge C_1$. $\mathit{cdf}_1$ and $\mathit{cdf}_2$ are the cumulative execution time distribution functions under configuration plan $C_1$ and $C_2$, respectively.}
\label{fig:cdf}       
\end{figure}

A hybrid instance configuration is represented as a vector
of both spot and on-demand instance types, as described in Section~\ref{subsec:term}.
The last dimension in the vector is the on-demand instance type obtained from
the $A^\star$-based instance configuration step.
Starting from the initial hybrid configuration to be only one on-demand type,
we repeatedly add spot instances in front of the hybrid configuration to
find the optimal hybrid configuration.
Refining a hybrid instance configuration $C_1$ of task $T$ to a hybrid instance configuration $C_2$,
we need to determine whether $C_2$ is better than $C_1$ in terms of execution time distributions.
Particularly, we define $C_2 \ge C_1$ if for $\forall t$, we have
$\int_0^t {P_{T,C_2}(time=x)}\, \mathrm{d}x \ge \int_0^t {P_{T,C_1}(time=x)}\, \mathrm{d}x$, where
$P_{T,C_1}$ and $P_{T,C_2}$ are the PDFs of task $T$ under hybrid instance configuration $C_1$
and $C_2$, respectively.
Figure~\ref{fig:cdf} illustrates this definition. The integrals are represented as CDFs (Cumulative distribution functions).

Ideally, we can add $n$ spot instances ($n$ is a predefined parameter).
A larger $n$ gives higher probability of
benefiting from the spot instances while a smaller $n$ gives
higher probability of meeting deadline requirement and reduces the optimization overhead.
In our experiments, we find that $n=2$ is sufficient for obtaining good
optimization results. A larger $n$ greatly increases the
optimization overhead with small improvement on the optimization results.

For each added spot instance in the hybrid instance
configuration, we need to decide its type and associated bidding price.
Due to price dynamics of spot instances, making the decision is non-trivial.
We search for the spot instance types
and the associated bidding prices as described in Algorithm~\ref{alg:spottune}.
To reduce the search space, we design a heuristic.
The added spot instance type should be at least as expensive
as the on-demand instance type in the hybrid configuration in order to
ensure the probabilistic deadline guarantee. For each spot instance type,
we search the bidding price in the range of $\mathit{[P_{min}, P_{max}]}$ using the search algorithm
described in Algorithm~\ref{alg:binary}. In our implementation,
$\mathit{P_{min}}$ is $0.001$ and $\mathit{P_{max}}$ equals to the price of the
on-demand instance of the same type.
The spot instance with the bidding price higher than $P_\mathit{max}$
does not contribute to monetary cost reduction.
If an bidding price
is found for a certain instance type, this instance type is assigned to the
hybrid instance configuration with the found bidding price. If no bidding price is suitable
for any spot instance type, the added spot instance type is assigned with
$-1$ indicating the spot instance is not added.

\begin{algorithm}
\begin{small}
\caption{Hybrid instance configuration refinement for a task $T$.} \label{alg:spottune}
\begin{algorithmic}[1]

\REQUIRE {
$\mathit{ondemand\_type}$: the on-demand type found in $A^\star$ instance configuration for task $T$\\
$n$: the dimensions of the hybrid instance configuration}

\STATE {$T$.configList[$n$] := $\mathit{ondemand\_type}$;}
\STATE {$T$.prices[$n$] := on-demand price of $\mathit{ondemand\_type}$;}
\FOR {$dim$ := 1 \TO $n-1$}
    \STATE {$T$.configList[$dim$] := $-1$;}
    \STATE {$T$.prices[$dim$] := $0$;}
\ENDFOR
\FOR {$dim$ := 1 \TO $n-1$}
    \FOR {$\mathit{spottype := ondemand\_type}$ \TO the most expensive instance type}
        \STATE {$\mathit{P_{max}}$ := the on-demand price of instance type $spottype$;}
        \STATE {$\mathit{P_b}$ := binary\_search($P_{min}$, $\mathit{P_{max}}$, $T$, $\mathit{spottype}$);}
        \IF {$\mathit{P_b} == -1$}
            \STATE {continue;}
        \ELSE
            \STATE {$T$.configList[$dim$] := $\mathit{spottype}$;}
            \STATE {$T$.prices[$dim$] := $\mathit{P_b}$;}
        \ENDIF
    \ENDFOR
\ENDFOR
\end{algorithmic}
\end{small}
\end{algorithm}

\begin{algorithm}
\begin{small}
\caption{Binary\_search function for task $T$.} \label{alg:binary}
\begin{algorithmic}[1]

\REQUIRE{
$\mathit{P_{low}}$: the lowest bidding price searched\\
$\mathit{P_{high}}$: the highest bidding price searched \\
$\mathit{spottype}$: the spot instance type }

\IF {$\mathit{P_{low} > P_{high}}$}
    \STATE Return -1;
\ENDIF
\STATE $\mathit{P_{mid}}$ := $\mathit{(P_{low} + P_{high})/2}$;
\STATE $\mathit{spotcost}$ := $\mathit{estimate\_cost}$(hybrid config with bidding price $\mathit{P_{mid}}$);
\STATE $\mathit{ondemandcost}$ := $\mathit{estimate\_cost}$(on-demand config);
\IF {$\mathit{spotcost > ondemandcost}$}
    \STATE {Return binary\_search($\mathit{P_{low}, P_{mid}, spottype}$);}
\ELSE
    \STATE $\mathit{satisfied}$ := $\mathit{estimate\_performance}$(hybrid config, $T$);
    \IF {\NOT $\mathit{satisfied}$}
        \STATE {Return binary\_search($\mathit{P_{mid}, P_{high}, spottype}$);}
    \ENDIF
    \STATE Return $\mathit{P_{mid}}$;
\ENDIF
\end{algorithmic}
\end{small}
\end{algorithm}

In Algorithm~\ref{alg:binary}, the bidding price is searched using a binary search algorithm.
We have the following two considerations: first, adding the spot instance should not violate
the probabilistic deadline guarantee; second, the estimated cost of the refined hybrid configuration
should be less than that before the refinement. If both considerations
are satisfied for a certain bidding price, we return this price
to the hybrid instance configuration.

{\bf Probabilistic deadline guarantee consideration.}
To ensure the probabilistic deadline guarantee, one way is to calculate the
probabilistic distribution of the entire workflow execution time and then to
decide whether the required percentile
in the probabilistic distribution is smaller than deadline.
However, this calculation requires large overhead. We implement
this process in the Oracle algorithm presented in Section~\ref{sec:eval} while
in Dyna, we propose a light-weight localized heuristic to reduce the overhead.
As the on-demand configuration found in the $\mathit{A^\star}$-based instance configuration
step has already ensured the probabilistic deadline requirement,
we only need to make sure that the hybrid configuration of each task $\mathit{C_{hybrid}}$
satisfies $\mathit{C_{hybrid}\ge C_{ondemand}}$ where $\mathit{C_{ondemand}}$ is the on-demand only configuration.
If this requirement is not satisfied,
it means the current bidding price is too low and
we continue the search of the bidding price in the higher half
of the current search range.

Since a spot instance may fail at any time, we define a probabilistic
function $\mathit{ffp(t,p)}$ to calculate
the the probability of a spot instance fails at time $t$ for
the first time when the bidding price is set to $p$.
Existing studies have demonstrated that the spot prices can be
predicted using statistics models~\cite{ModelSpot} or reverse
engineering~\cite{Deconstructing}.
We use the recent spot price history as a prediction of the real spot price
for $\mathit{ffp(t,p)}$ to calculate the failing probability.
We obtain that function with a Monte-Carlo based approach.
Starting from a random point in the price history, if the price history
becomes larger than $p$ at time $\mathit{t}$ for the first time,
we add one to the counter $\mathit{count}$.
We repeat this process for $\mathit{NUM}$ times ($\mathit{NUM}$
is sufficiently large) and return $\mathit{\frac{count}{NUM}}$ as the failing probability.

In order to decide whether a refined hybrid instance configuration is better,
we first discuss how to estimate the overall execution time distribution of a task
given a hybrid instance configuration.
Assume a hybrid instance configuration of task $T$ is
$\mathit{C_{hybrid}}$ = $<$ ($\mathit{type_1}$, $\mathit{P_b}$, $\mathit{isSpot}$),
($\mathit{type_2}$, $\mathit{P_o}$, $\mathit{isOndemand}$) $>$.
Assume the probabilistic distributions of task $T$ on the spot instance of $\mathit{type_1}$
is $\mathit{P_{T,type_1}}$ and on the on-demand instance of $\mathit{type_2}$ is $\mathit{P_{T,type_2}}$.
The overall execution time of task $T$ under $\mathit{C_{hybrid}}$
can be divided into two cases. If the task successfully finishes on the spot instance,
the overall execution time equals to the execution time of task $T$ on the spot instance $t_s$ with
the following probability.
\begin{align}
\mathit{P_{T,C_{hybrid}}(time=t_s)} &= \mathit{P_{T,type_1}(time=t_s)}\times \nonumber \\
&(1-\int_0^{t_s} \mathit{ffp(x,P_b)}\, \mathrm{d}x)
\end{align}
Otherwise, the overall execution time equals to the time task $T$ has run on the spot
instance before it fails, $t_f$, plus the execution time of task $T$ on the on-demand instance $t_o$,
with the following probability.
\begin{align}
\mathit{P_{T,C_{hybrid}}(time=t_f+t_o)} &= \mathit{P_{T,type_1}(time=t_f)}\times \nonumber \\ &\mathit{P_{T,type_2}(time=t_o)}\times \nonumber \\
&\mathit{ffp(t_f,P_b)} (t_f\le t_s)
\end{align}

After obtaining the execution time distribution of a task under the hybrid
instance configuration $\mathit{C_{hybrid}}$, we compare it with the on-demand configuration
$\mathit{C_{ondemand}}$ using the definition
shown in Figure~\ref{fig:cdf}. If $\mathit{C_{hybrid}\ge C_{ondemand}}$ is satisfied,
we accept the refined hybrid instance configuration for the task.
We use this heuristic to localize the
computation of the execution time distribution of the entire workflow
to each task and greatly reduce the optimization overhead.

{\bf Monetary cost consideration.}
If the expected monetary cost of the refined configuration is higher than that
before the refinement, it means the bidding price
is set too high. Thus we continue the search of the bidding price in the
lower half of the search range.
We estimate the cost of a hybrid configuration of a task as described
in Algorithm~\ref{alg:hybridcost}.
Similar to the analysis for performance estimation, we
obtain the execution time distribution of the
hybrid configuration and calculate the expected monetary cost
under the searched bidding price.
We compare the expected monetary cost of the
hybrid configurations before and after refinement,
and add the spot instance only if the refinement can reduce the expected monetary cost.

\begin{algorithm}
\begin{small}
\caption{Estimate monetary cost, $\mathcal{C}$, for a task with
hybrid instance configuration of $2$ instances} \label{alg:hybridcost}
\begin{algorithmic}[1]
\REQUIRE Hybrid instance configuration of the task $<$$\mathit{type_1}$, $\mathit{P_b}$,
$\mathit{isSpot}$ $>$ and $<$$\mathit{type_2}$, $\mathit{P_o}$,
$\mathit{isOndemand}$ $>$;\\
Execution time samples of the task on $type_1$ instance: $stime_i$ ($1\le i \le N$);\\
Execution time samples of the task on $type_2$ instance: $otime_i$ ($1\le i \le N$);

\STATE $\mathcal{C}=0$;
\FOR {$\mathit{i=1;~i\le N;~i=i+1}$}
    \STATE $p=0$;
    \FOR {$\mathit{t=0;~t<stime_i;~t=t+step}$}
        \STATE $\mathit{p=p+ffp(t,P_b)}$;
    \ENDFOR
    \STATE {$\mathit{\mathcal{C}=\mathcal{C}+P_b\times stime_i+p\times P_o\times otime_i}$};
\ENDFOR
\STATE $\mathcal{C} = \frac{\mathcal{C}}{N}$;
\STATE Return $\mathcal{C}$;
\end{algorithmic}
\end{small}
\end{algorithm}

\subsection{Dynamic Optimizations}

\begin{figure}
\centering
  \includegraphics[width=0.3\textwidth]{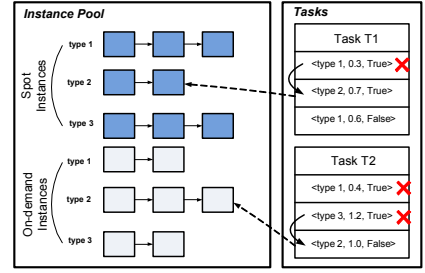}
\centering
  \caption{An example of hybrid execution}
\label{sec3:fig:2}       
\end{figure}

When a task's all preceding tasks in the
workflow have finished execution, the task becomes ready and can be
scheduled to execute on an instance according to the instance
configuration determined from the static optimization.
In the dynamic optimization stage,
we adopt the consolidation and scaling technique
and implement online instance reuse to further optimize execution cost.
These techniques have been widely used in the previous studies~\cite{SC11,Malawski:2012:CDP:2388996.2389026}.
Still, we need to extend them with the consideration of hybrid instance
configuration. In the following, we briefly introduce those techniques,
with the special attention to the difference brought by supporting
hybrid instance configurations.

{\bf Consolidation and Scaling.}
Consolidating different
instances could reduce the new instance acquisition time as well as
the instance-hour billing overhead. For each instance in the pool,
we record the instance's remaining partial hours.

Due to the difference between spot and on-demand instances, virtual
machine consolidation among them needs special care. If the
remaining time of the spot (resp. on-demand) instance is long enough
for a ready task's execution, and the task
is going to acquire a spot (resp. on-demand) instance of the same
type, the task will be assigned to the instance.
The consolidation
between spot and on-demand instances is possible only for the case
whereby spot instance request is consolidated to an on-demand
instance, due to the success guarantee.

{\bf Online Instance Reuse.}
At runtime, we maintain a pool
of running instances, organized in lists according to different
instance types.
The spot and on-demand instances with the same instance type are organized in
separated lists. During runtime, an instance request is processed at the instance
starting time by first looking into the instance pool for an
idle instance of the requested type. If such an instance is
found, it will be selected for workflow execution. Otherwise,
a new instance of the requested type is acquired from the
IaaS cloud. Thus, the instances started during workflows
execution can be properly reused and their utilizations are
improved. Additionally, if we can reuse the instances, the
instance acquisition time is eliminated.

Figure~\ref{sec3:fig:2} shows
an example of the pool of current spot and on-demand instances
as well as two tasks named {\sf T1} and {\sf T2} to be executed.
Initially, task {\sf T1} is assigned to a spot
instance with type $\mathit{type}_1$. But the execution of task {\sf T1}
fails due to out-of-bid event and restarts on a
$\mathit{type}_2$ spot instance. If there is no $\mathit{type}_2$
spot instance available in the instance pool, the task will wait for
the setup of a new $\mathit{type}_2$ spot instance which will be bid
with the price indicated in the task's hybrid instance configuration. At the
meantime, as {\sf T2} has failed on both $\mathit{type}_1$ and
$\mathit{type}_3$ spot instances, it will execute on a $\mathit{type}_2$
on-demand instance. Similarly, if there is no $\mathit{type}_2$
on-demand instance in the pool, a new $\mathit{type}_2$ on-demand
instance is requested first and added into the instance pool.

\section{Evaluation}\label{sec:eval}
In this section, we present the evaluation results of the proposed
approach on Amazon EC2.
\subsection{Experimental Setup}

We have two sets of experiments on real cloud environments: firstly
calibrating the cloud dynamics from Amazon EC2 as the input of our
optimization framework and secondly
running real-world scientific workflows on Amazon EC2 with
the compared algorithms for evaluation.

{\bf Calibration.} We measure the performance of CPU, I/O and
network for four frequently used instance types, namely m1.small,
m1.medium, m1.large and m1.xlarge. We find that CPU performance
is rather stable, which is consistent with the previous
studies~\cite{Runtimemeasure}. Thus, we focus on the calibration for
I/O and network performance. In particular, we repeat the
performance measurement on each kind of instance for
$1,0000$ times. When an instance has been acquired for a
full hour, it will be released and a new instance of the same type
will be created to continue the measurement. The measurement results
are used to model the probabilistic distributions of I/O and network
performance.

We measure both sequential and random I/O performance for local
disks. The sequential I/O reads performance is measured with
\emph{hdparm}. The random I/O performance is measured by generating random I/O reads of 512 bytes
each. Reads and writes have similar performance results, and we do
not distinguish them in this study.

We measure the uploading and downloading bandwidth between different
types of instances and Amazon S3. The bandwidth is measured from
uploading and downloading a file to/from S3. The file size is set to
$\mathit{8 MB}$. We also measured the network bandwidth between two
instances using Iperf~\cite{iperf}. We find
that the network bandwidth between instances of different types is generally
lower than that between instances of the same type and S3.

\begin{figure}
\centerline{
  \includegraphics[width=0.4\textwidth]{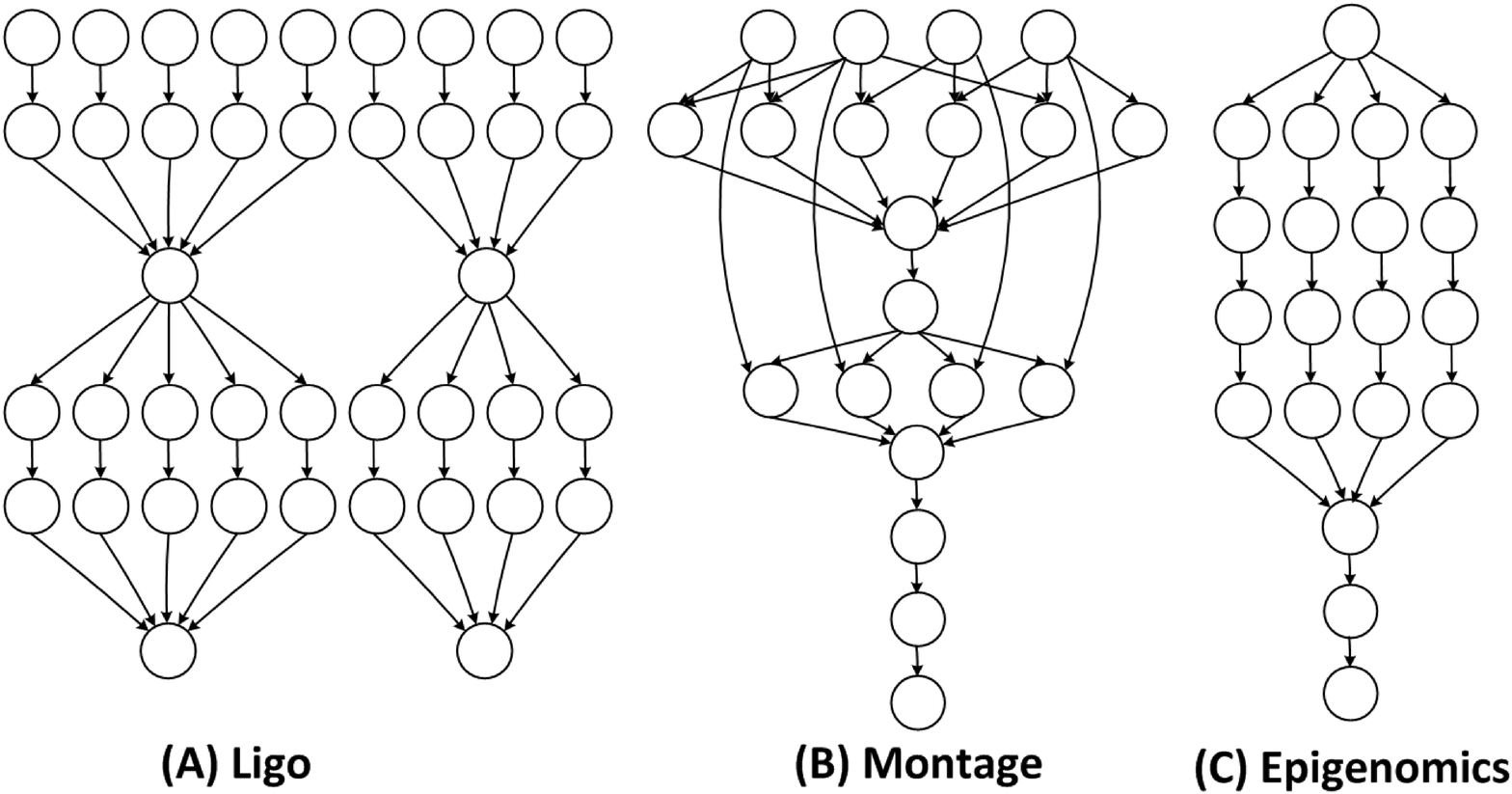}}
 \caption{Workflow structures: Ligo, Montage and Epigenomics.}
\label{sec4workflow}       
\end{figure}

{\bf Workload.}
There have been some works on characterizing the performance behaviours of scientific
workflows~\cite{characterization1,characterization2}. In this paper,
we consider three common workflow structures, namely \emph{Ligo}, \emph{Montage} and \emph{Epigenomics}.
Ligo (Laser Interferometer Gravitational Wave Observatory) is an application used to
detect gravitational-wave.
Montage is an astronomical application widely used as a Grid and
parallel computing benchmark.
Epigenomics is a data processing pipeline of various genome sequencing operations.
As shown in Figure~\ref{sec4workflow}, the three workflows have different structures and parallelism.
They also have different requirements on computation resources. For example, Montage is I/O-intensive
workload, Ligo is memory-intensive workload and Epigenomics is CPU-intensive. Thus, with the three
workflows, we are able to examine the effectiveness of our proposed algorithm on different workloads.
These workflows have been characterized in details by Juve et al~\cite{characterization1} and readers can refer to their work for more detailed information.

{\bf Comparisons.}
In order to evaluate the effectiveness of the proposed techniques in Dyna, we have implemented the following algorithms on Amazon EC2.
\begin{itemize}
  \item
  {\bf Static.} This approach is the same as the previous study
  in~\cite{SC11} which only adopts on-demand instances. We adopt it as the state-of-the-art comparison.
  For a fair comparison, we set the workflow deadline according to the QoS setting used in Dyna. For example, if user requires $90\%$ of probabilistic deadline, the deterministic deadline used in the experiment is set to the 90-th percentile of its execution time distribution.
  \item
  {\bf Dyna-NS.} This approach is the same as Dyna except that Dyna-NS does not use any spot instances. The comparison between Dyna and Dyna-NS is to assess the impact of spot instances.
  \item
  {\bf SpotOnly.} This approach adopts only spot instances during execution. It first utilizes the $A^\star$-based instance configuration approach to decide the instance type for each task in the workflow. Then we set the bidding price of each task to be very high (in our studies, we set it to be \$1, 000) in order to guarantee the required deadline meeting rate.
  \item
  {\bf Oracle.} We implement the Oracle method to assess the tradeoff between the optimization overhead and the
  effectiveness of the optimizations in Dyna. Oracle is different from Dyna in the following key designs. In the binary search method, which is used to search bidding prices for spot instances in the hybrid configurations of tasks, Oracle does not adopt the localized heuristic for percentile calculation as Dyna. On the contrary, for each searched bidding price of a task, it calculates the overall execution time distribution of the workflow to decide if the searched bidding price for the task satisfies the probabilistic deadline requirement. This is an offline approach, since the time overhead of getting the solution in Oracle is prohibitively high.
\end{itemize}

We have adopted DAGMan~\cite{dagman} to manage task dependencies and added
Condor~\cite{condor} to the Amazon Machine Image (AMI) to manage
task execution and instance acquisition.

We acquire the instances from the US East region. The hourly costs of
the on-demand instance for the four instance types used in workflow execution are shown in
Table~\ref{tb:spot}. Those four instances have also been used in the previous
studies~\cite{INRIA}.
As for the instance acquisition time ($\mathit{lag}$), our
experiments show that each on-demand instance
acquisition costs 2 minutes and spot instance acquisition
costs 7 minutes on average. This is consistent with the
existing studies by Mao et al.~\cite{DBLP:conf/IEEEcloud/MaoH12}.

The deadline of workflows is an important factor for the candidate space of
determining the instance configuration. There are two
deadline settings with particular interests: $D_{min}$ and
$D_{max}$, the expected execution time of all the tasks in the
critical path of the workflow all on the m1.xlarge and m1.small instances,
respectively. By default, we set the deadline to be
$\frac{D_{min}+D_{max}}{2}$.

We assume there are many workflows submitted by the users to the WaaS provider.
In each experiment, we submit 100 jobs of the same workflow structure
to the cloud. We assume the job arrival conforms to a Poisson distribution. The
parameter $\lambda$ in the Poisson distribution affects the chance
for virtual machine reuse. By default, we set $\lambda$ as 0.1.

As for metrics, we study the average monetary cost and elapsed time
for a workflow. \emph{All the metrics are normalized to those of Static.}
Given the probabilistic deadline
requirement, we run the compared algorithms multiple times on the cloud
and record their monetary cost and execution time.
We consider monetary cost as the main matric for
comparing the optimization effectiveness of different scheduling algorithms
when they all satisfy the QoS requirements.
By default, we set the required deadline hit rate as 96\%.
By default, we present the results obtained when all parameters are set to their default setting.
In Section~\ref{subsec:sensit},
we experimentally study the impact of different parameters with sensitivity studies.

\subsection{Cloud Dynamics}
\label{subsec:calib}
\begin{figure}
\centering
  \includegraphics[width= 0.3\textwidth]{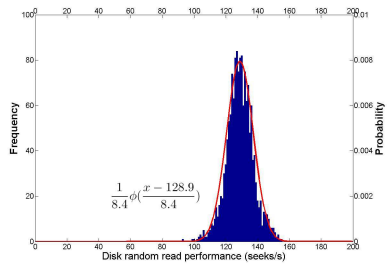}
\centering
  \caption{The histogram and probabilistic distribution of random I/O performance on Medium instances}
\label{fig:io_perf}       
\end{figure}

\begin{figure}
\centering
  \includegraphics[width= 0.3\textwidth]{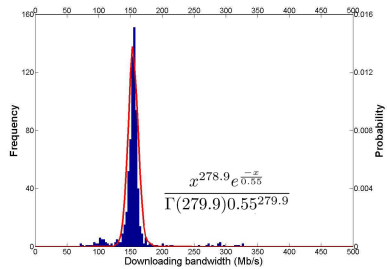}
\centering
  \caption{The histogram and probability distribution of downloading bandwidth between Medium instances and S3 storage}
\label{fig:net_perf}       
\end{figure}
In this subsection, we present the performance dynamics observed on Amazon
EC2. The price dynamics have been presented in Table~\ref{tb:spot} of
Section~\ref{sec:problem}.

Figure~\ref{fig:io_perf} and Figure~\ref{fig:net_perf} show the
measurements of random I/O performance and downloading network performance from Amazon S3 of
Medium instances. We have observed similar results on other instance
types. We make the following observations.

First, both I/O and network performances can be modeled with normal
or Gamma distributions. We verify the distributions with null
hypothesis, and find that (1) the sequential I/O performance and
uploading and downloading network bandwidth from/to S3 of the four
instance types follow Gamma distribution; (2) the random I/O
performance distributions on the four instance types follow normal distribution.
The parameters of these distributions are presented in
Tables~\ref{table:io_perf} and \ref{table:net_perf}.

Second, the I/O and network performance of the same instance type
varies significantly, especially for m1.small and m1.medium instances.
This can be observed from the $\theta$ parameter of Gamma
distributions or the $\sigma$ parameter of normal distributions in
Tables~\ref{table:io_perf} and \ref{table:net_perf}. Additionally,
random I/O performance varies more significantly than sequential I/O
performance on the same instance type. The coefficient of variance
of sequential and random I/O performance on m1.small are $9\%$ and
$33\%$, respectively.

Third, the performance between different instance types also differ
greatly from each other. This can be observed from the $k\cdot
\theta$ parameter (the expected value) of Gamma distributions or the
$\mu$ parameter of normal distributions in
Tables~\ref{table:io_perf} and \ref{table:net_perf}.

 \begin{table}
  \centering
  \caption{Parameters of I/O performance distributions}\label{table:io_perf}
  \begin{tabular}{|c|c|c|}
    \hline
    Instance type& Sequential I/O (Gamma)& Random I/O (Normal)\\\hline
    m1.small&$k = 129.3, \theta = 0.79$&$\mu = 150.3, \sigma = 50.0$ \\\hline
    m1.medium&$k = 127.1, \theta = 0.80$&$\mu = 128.9, \sigma = 8.4$ \\\hline
    m1.large&$k = 376.6, \theta = 0.28$&$\mu = 172.9, \sigma = 34.8$ \\\hline
    m1.xlarge&$k = 408.1, \theta = 0.26$&$\mu = 1034.0, \sigma = 146.4$ \\\hline
\end{tabular}
\end{table}

 \begin{table}
  \centering
  \caption{Gamma distribution parameters on bandwidth between an instance and S3}\label{table:net_perf}
  \begin{tabular}{|c|p{3cm}|p{3cm}|}
    \hline
    Instance type& Uploading bandwidth & Downloading bandwidth \\\hline
    m1.small&$k = 107.3, \theta = 0.55$&$k= 51.8, \theta = 1.8$ \\\hline
    m1.medium&$k = 421.1, \theta = 0.27$&$k= 279.9, \theta = 0.55$ \\\hline
    m1.large&$k = 571.4, \theta = 0.22$&$k= 6187.7, \theta = 0.44$ \\\hline
    m1.xlarge&$k = 420.3, \theta = 0.29$&$k= 15313.4, \theta = 0.23$ \\\hline
\end{tabular}
\end{table}

\subsection{Overall Comparison}

\begin{figure}
  \centering
  \includegraphics[width= 0.3\textwidth]{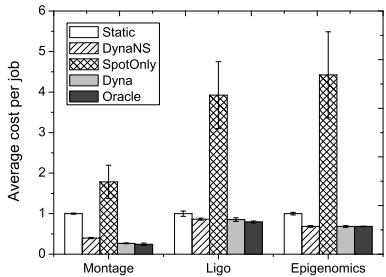}
  \caption{The average monetary cost optimization results of compared algorithms on Montage, Ligo and Epigenomics workflows in Amazon EC2. }\label{fig:overallcost}
\end{figure}
\begin{figure}
  \centering
  \includegraphics[width=0.3\textwidth]{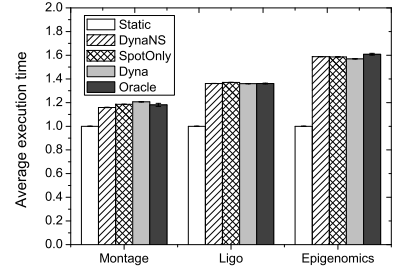}
  \caption{The average execution time optimization results of compared algorithms on Montage, Ligo and Epigenomics workflows in Amazon EC2.}\label{fig:overalltime}
\end{figure}

\begin{figure}
\centering
\includegraphics[width=0.3\textwidth]{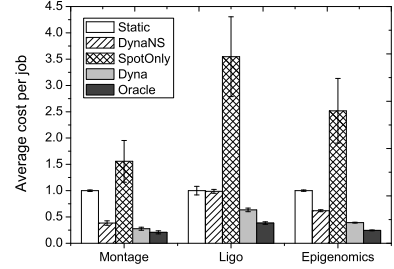}
\caption{The average monetary cost results of compared algorithms on Montage, Ligo and Epigenomics workflows when the probabilistic deadline guarantee is 90\%.}\label{fig:overallcost90}
\end{figure}

\begin{table}
\centering\begin{small}
\caption{Obtained deadline hit rates by the compared algorithms on Montage, Ligo and Epigenomics workflows when the probabilistic deadline guarantee is 96\%.}\label{tb:overallhitrate}
\begin{tabular}{|c|c|c|c|c|c|}
\hline
&Static&DynaNS&SpotOnly&Dyna&Oracle \\\hline
Montage&100\%&96.5\%&96.7\%&96.7\%&96.6\% \\\hline
Ligo&100\%&96.2\%&96.8\%&96.8\%&96.6\% \\\hline
Epigenomics&100\%&96.8\%&96.7\%&96.8\%&96.7\% \\\hline
\end{tabular}
\end{small}
\end{table}

\begin{table}
\centering\begin{small}
\caption{Optimization overhead of the compared algorithms on Montage, Ligo and Epigenomics workflows (seconds).}\label{tb:overhead}
\begin{tabular}{|c|c|c|c|c|c|}
\hline
&Static&DynaNS&SpotOnly&Dyna&Oracle \\\hline
Montage&1&153&153&163&2997 \\\hline
Ligo&1&236s&236&244&10452\\\hline
Epigenomics&1&166&166&175&2722 \\\hline
\end{tabular}\end{small}
\end{table}

\begin{figure}
  \centering
  \begin{minipage}{0.45\linewidth}
  \subfigure[Histogram of the spot price of m1.small instance type]{
  \includegraphics[width= 0.95\textwidth]{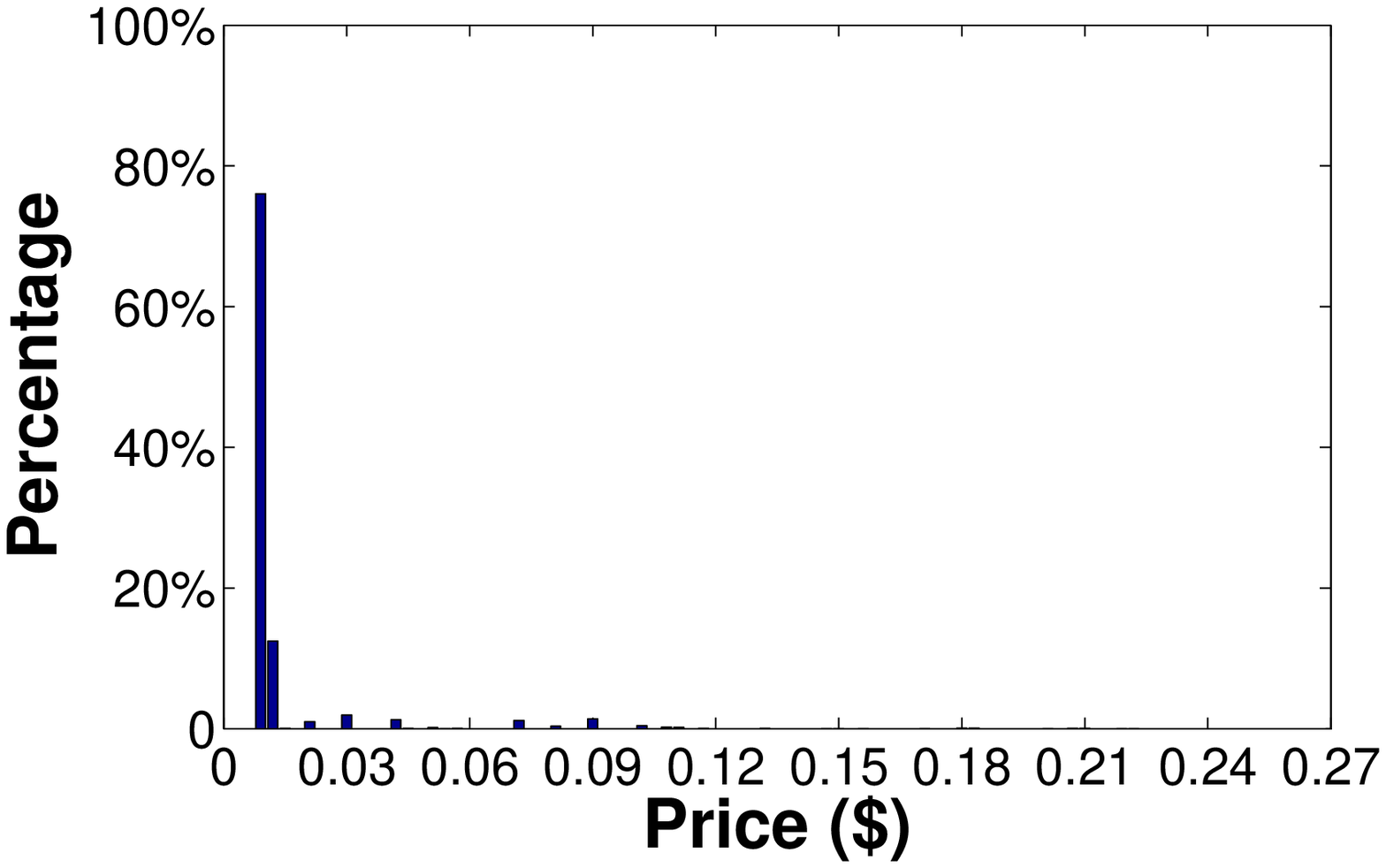}}
  \end{minipage}
  \begin{minipage}{0.45\linewidth}
  \subfigure[Histogram of the spot price of m1.xlarge instance type]{
  \includegraphics[width=0.95\textwidth]{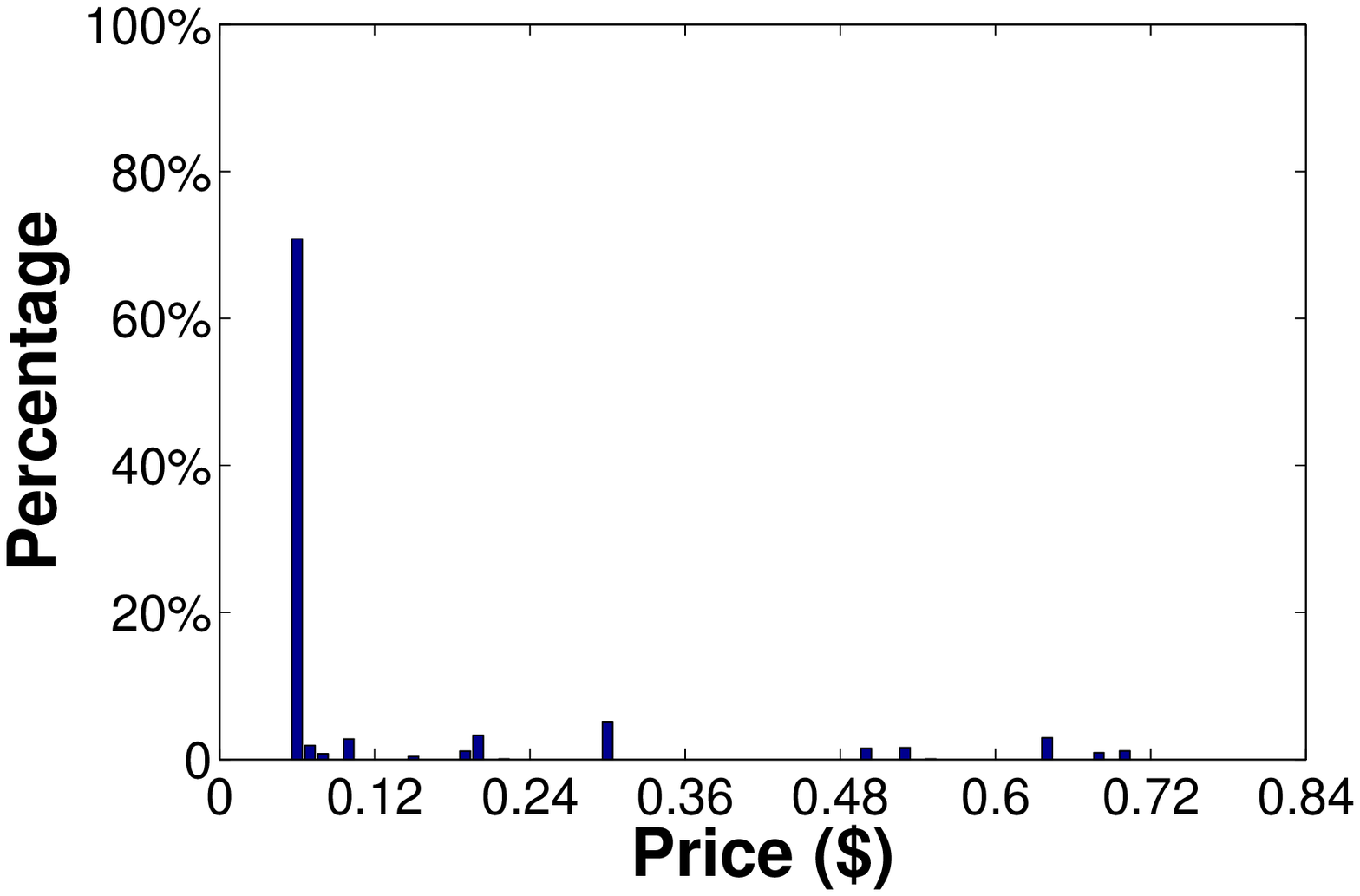}}
  \end{minipage}
  \caption{Histogram of the spot price history in August 2013, US East Region of Amazon EC2.}\label{fig:spotprice}
\end{figure}

In this sub-section, we present the overall comparison results of
Dyna and the other compared algorithms on Amazon EC2 under the default settings.
Sensitivity studies are presented in Section~\ref{subsec:sensit}.
Table~\ref{tb:overallhitrate} shows the obtained deadline hit rates.
Note that we have used the calibrations from Section~\ref{subsec:calib}
as input to the Dyna optimizations.
The results demonstrate the capability of our designed
techniques on accurately satisfying the required probabilistic deadline guarantees.
Although Static can guarantee the probabilistic deadline requirement
with a higher deadline hit rate, it causes a much higher
extra monetary cost of the WaaS provider, as we will demonstrate later in this subsection.

Figure~\ref{fig:overallcost} shows the average monetary cost per job
results of Static, DynaNS, SpotOnly, Dyna and Oracle
methods on the Montage, Ligo and Epigenomics workloads.
The standard errors of the monetary cost results of Static, DynaNS, SpotOnly, Dyna and
Oracle are 0.01--0.06, 0.02--0.04, 0.40--0.76, 0.01--0.03 and 0.01--0.03,
respectively, on the tested workloads.
Overall, Dyna obtains the smallest monetary cost among
the online approaches in all three workloads, saving more
monetary cost than Static, DynaNS and SpotOnly by 15--73\%, 1--33\% and 78\%--85\%,
respectively. We make the following observations.

First, DynaNS obtains smaller monetary cost than Static, because
the proposed $A^\star$ configuration search technique is capable of finding cheaper instance
configurations and is suitable for different structures
of workflows. This also shows that performing deadline assignment before instance configuration
in the Static algorithm reduces the optimization effectiveness.

Second, Dyna obtains smaller
monetary cost than DynaNS, meaning that the hybrid configuration with spot instances is effective
on reducing monetary cost. As the probabilistic deadline guarantee set lower, the monetary cost
saved by Dyna over DynaNS gets higher. Figure~\ref{fig:overallcost90} shows the monetary cost results
of the compared algorithms when the probabilistic deadline guarantee is set to 90\%. In this setting,
Dyna saves more monetary cost than DynaNS by 28--37\%.

Third, SpotOnly obtains the highest monetary cost among all the compared algorithms. This is due to the dynamic characteristic of spot price. Figure~\ref{fig:spotprice} shows the histogram of the spot price during the month of the experiments. Although the spot price is lower than the on-demand price of the same type in most of the time, it can be very high compared to on-demand price at some time. As shown in Table~\ref{tb:spot}, the highest spot price for a m1.small instance in August 2013 is \$10 which is more than 160 times higher than the on-demand price. The relative monetary cost of SpotOnly over the other compared algorithms is especially higher on Ligo because the average execution time of Ligo is longer than the other workflows.
Nevertheless, this observation depends on the fluctuation of spot price. The results on comparing SpotOnly and Dyna can be different if we run the experiments at other times. We
study the sensitivity of Dyna and SpotOnly to spot price with another spot price history in
Section~\ref{subsec:sensit}.

Figure~\ref{fig:overalltime} shows the average execution time of a workflow
of Static, DynaNS, SpotOnly, Dyna and Oracle
methods on the Montage, Ligo and Epigenomics workloads in Amazon EC2.
The standard errors of the execution time results of the compared algorithms are
small, all in 0.004--0.01 on the tested workloads.
Static has the smallest average execution time. This is because Static configures each task in workflows
with better and more expensive instance types. The average execution times of SpotOnly, Dyna and Oracle are similar. This is because the three algorithms all use the proposed static optimization to configure each task in the workflow with a certain instance type. Also, the careful selection of bidding price for each task in the workflow in Dyna and high bidding prices in SpotOnly diminishes the out-of-bid events during execution. On the other hand, we can also see that the bidding price searched during static optimization is able to diminish the out-of-bid events.

Finally, we analyze the optimization overhead of the compared algorithms.
The optimization overhead results are shown in Table~\ref{tb:overhead}.
Note that, for workflows with the same structure and profile,
our framework only need to do the optimization once.
Although Oracle obtains smaller monetary cost than Dyna, the optimization
overhead of Oracle is 16--44 times as high as that of Dyna.
This shows that Dyna is able to find optimization results close to the optimal results in much shorter time.
Due to the large optimization overhead, in the rest of the experiments, we do not evaluate Oracle but
only compare Dyna with Static, DynaNS and SpotOnly.

\subsection{Sensitivity studies}
\label{subsec:sensit}
We have conducted sensitivity studies on different workflows. Since
we observed similar results across workflows, we focus on
Montage workflows in the following. In each
study, we vary one parameter at a time and keep other parameters in their default settings.

\begin{figure}
  \centering
  \begin{minipage}{0.45\linewidth}
  \subfigure[Average monetary cost per job]{
  \includegraphics[width= 0.95\textwidth]{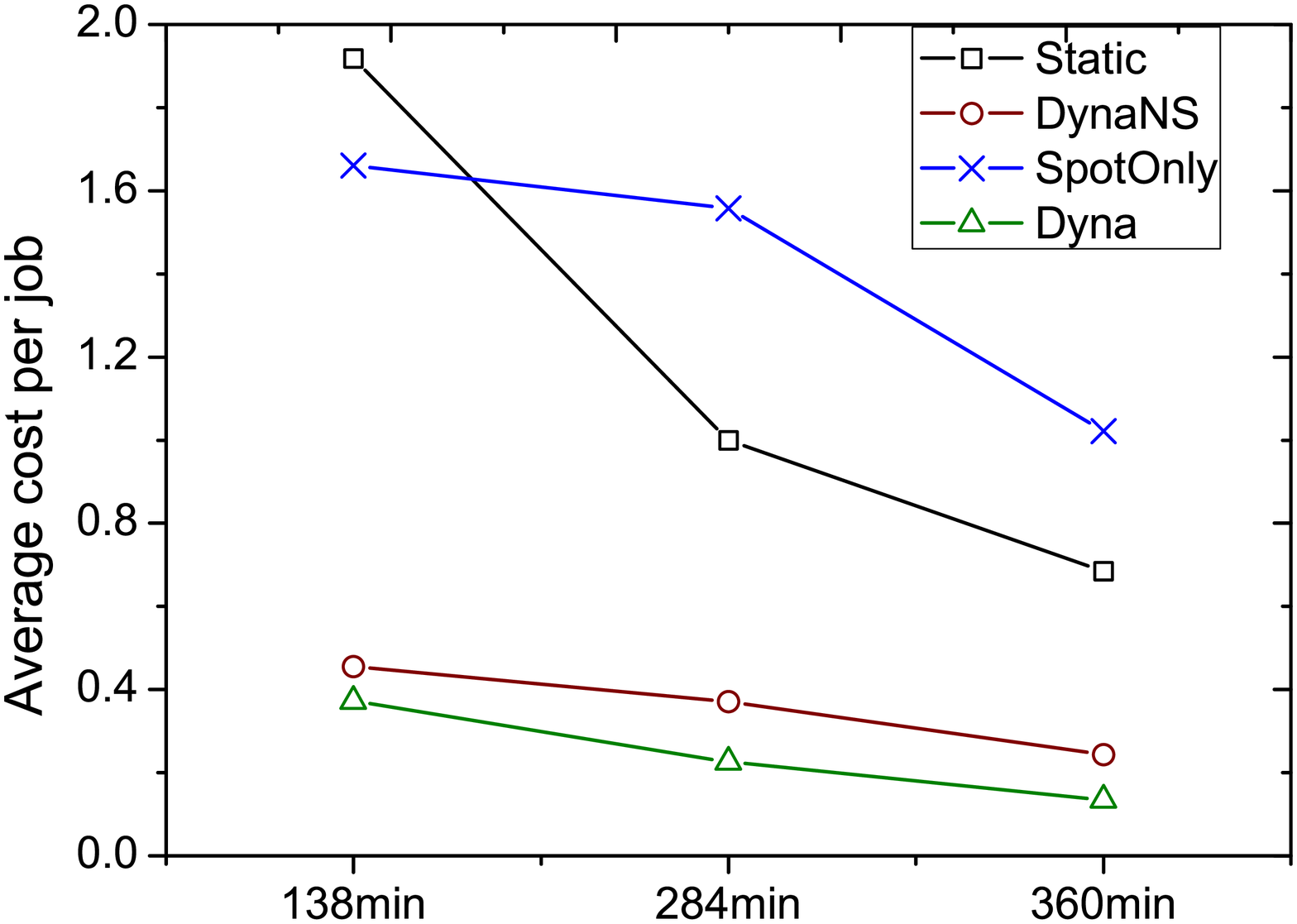}\label{fig:deadlinecost}}
  \end{minipage}
  \begin{minipage}{0.45\linewidth}
  \subfigure[Average execution time]{
  \includegraphics[width=0.95\textwidth]{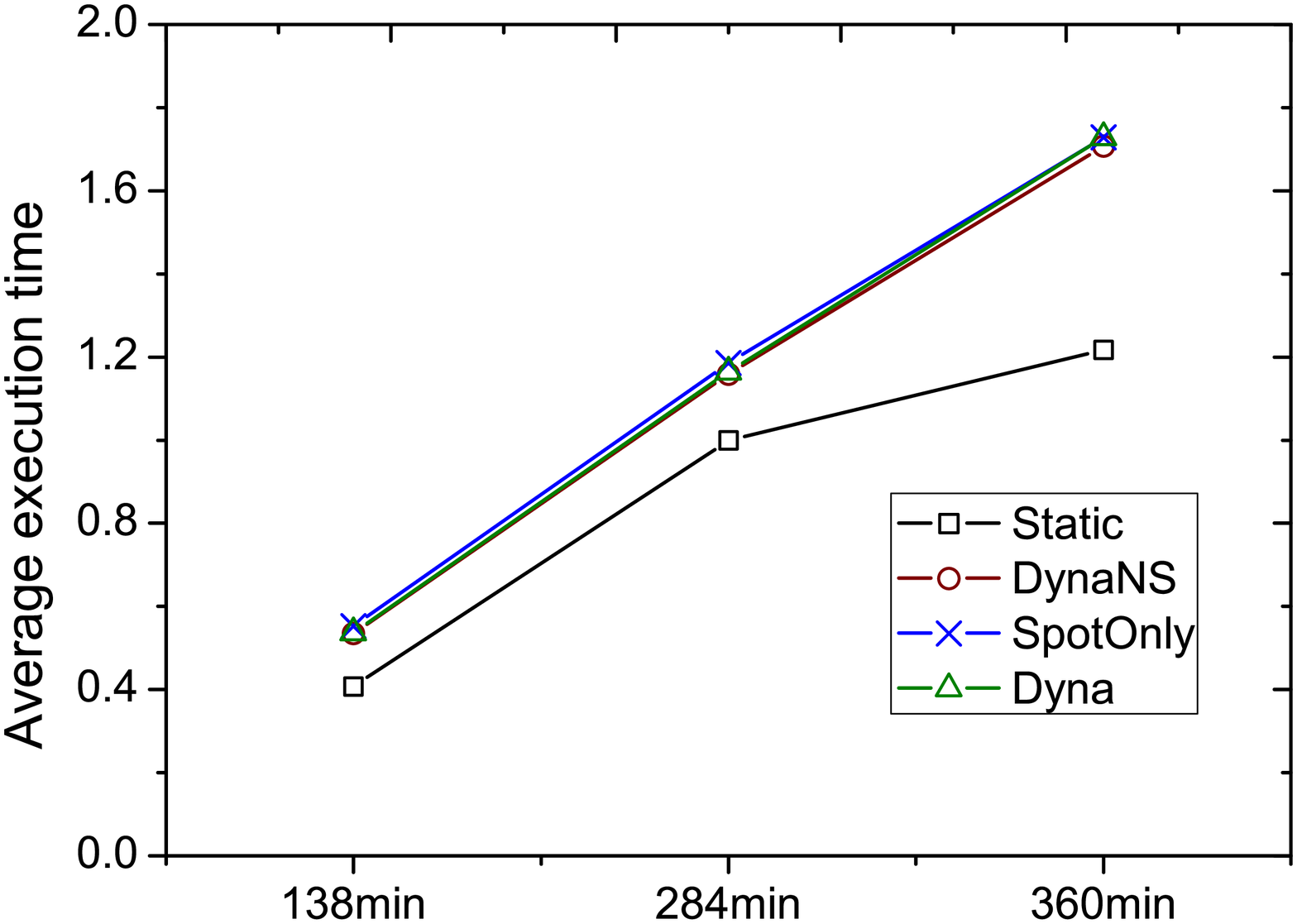}\label{fig:deadlinetime}}
  \end{minipage}
  \caption{The average monetary cost and average execution time results of sensitivity studies on deadline.}\label{fig:deadline}
\end{figure}

\begin{figure}
  \centering
  \begin{minipage}{0.45\linewidth}
  \subfigure[Breakdown when deadline is $1.5\times D_{min}$]{
  \includegraphics[width= 0.95\textwidth]{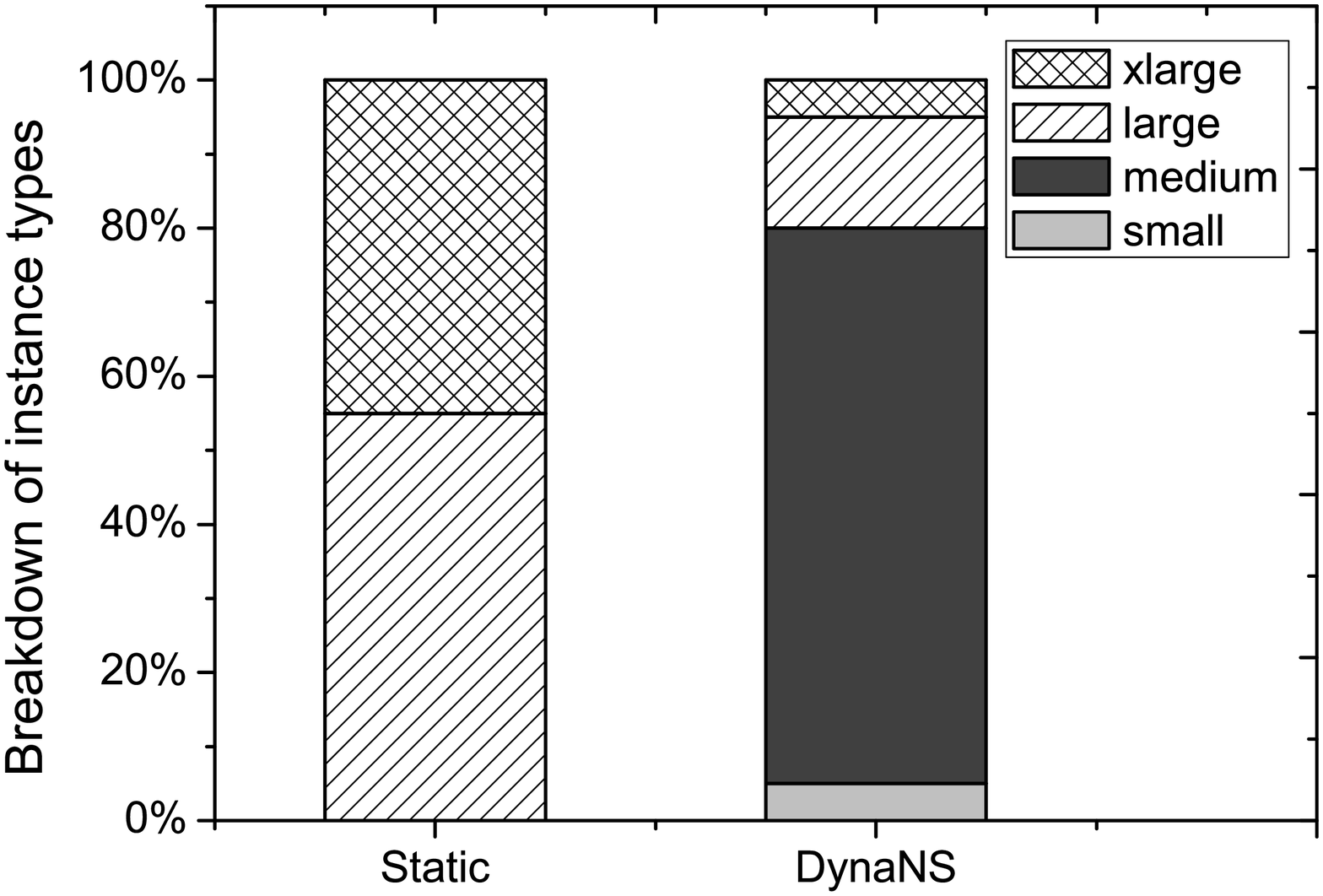}\label{fig:meetdlcost}}
  \end{minipage}
  \begin{minipage}{0.45\linewidth}
  \subfigure[Breakdown when deadline is $0.75\times D_{max}$]{
  \includegraphics[width=0.95\textwidth]{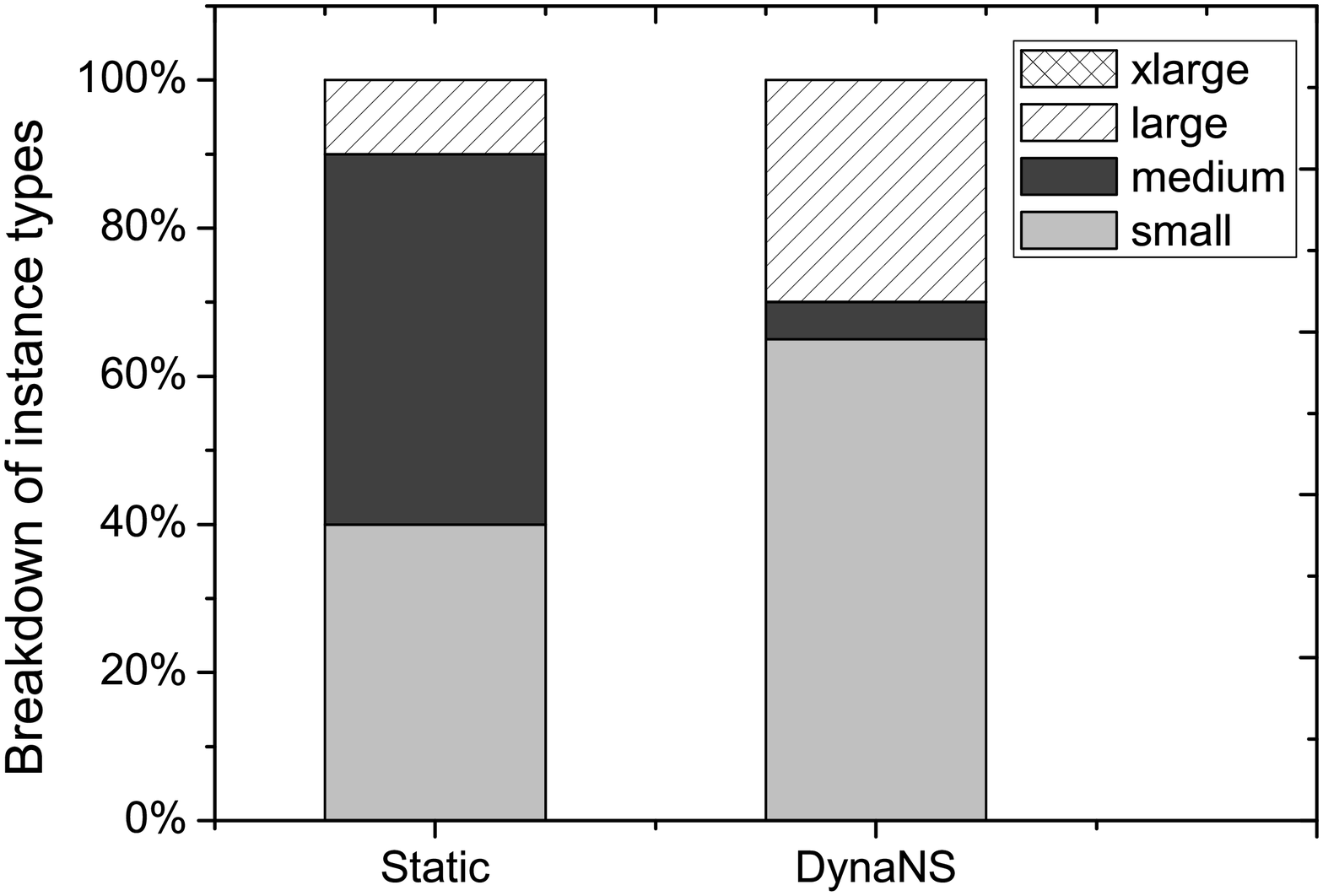}\label{fig:meetdltime}}
  \end{minipage}
  \caption{Breakdown of the instance types adopted by compared algorithms when the deadlines are $1.5\times D_{min}$ and $0.75\times D_{max}$.}\label{fig:deadlinebreakdown}
\end{figure}

{\bf Deadline.} Deadline is an important factor for determining
the instance configurations. We evaluate the compared algorithms under
deadline requirement varying from $1.5\times D_{min}$, $0.5\times
(D_{min}+D_{max})$ to $0.75\times D_{max}$, which are 138 minutes, 284 minutes
and 360 minutes on average for the Montage workflow, respectively. All results are normalized
to those of Static when deadline is $0.5\times(D_{min}+D_{max})$.
Figure~\ref{fig:deadline} shows the average monetary cost per job
and average execution time results. Dyna obtains the smallest
average monetary cost among the compared algorithms under all tested
deadline settings.
As the deadline gets loose, the monetary cost
is decreased since more cheaper instances (on-demand instances) are used
for execution. This trend does not apply to SpotOnly because
the spot price of the m1.medium instance can be lower than the m1.small instance
at some time. We have validated this phenomena with studying the spot price trace.
We further break down the
number of different types of on-demand and spot instances when the
deadlines are $\mathit{1.5\times D_{min}}$ and
$\mathit{0.75\times D_{max}}$ as shown in Figure~\ref{fig:deadlinebreakdown}.
The breakdown results of SpotOnly and Dyna are the same as DynaNS
because they all use the $A^\star$-based instance configuration method.
When the deadline is loose ($0.75\times D_{max}$), more cheap instances
are utilized.
Also, when under the same deadline, e.g., $\mathit{1.5\times D_{min}}$, DynaNS, SpotOnly and
Dyna utilize more cheap instances than Static, which
again shows our $A^\star$ approach is better than the existing heuristics~\cite{SC11}.

\begin{figure}
  \centering
  \begin{minipage}{0.45\linewidth}
  \subfigure[Average monetary cost per job]{
  \includegraphics[width= 0.95\textwidth]{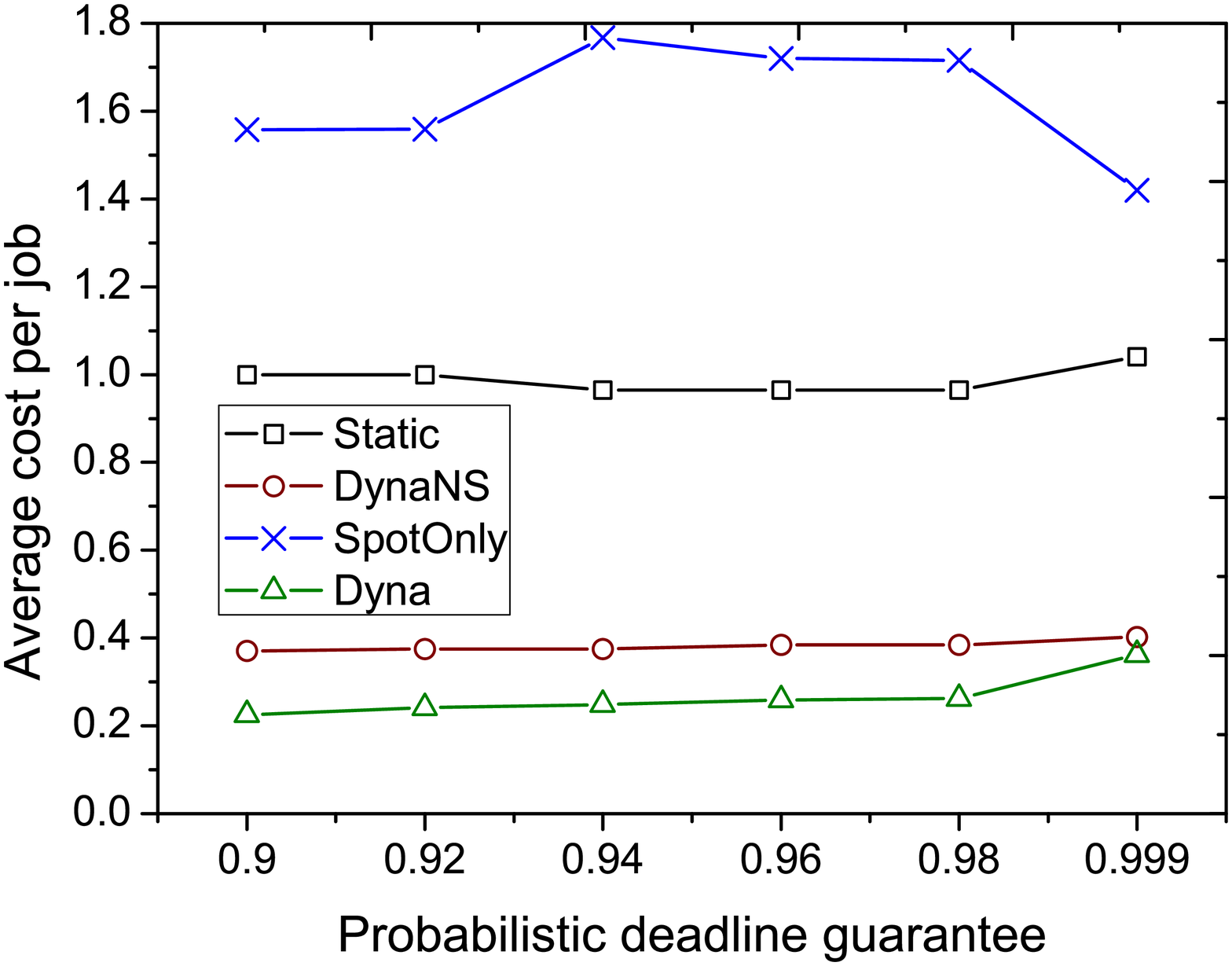}\label{fig:meetdlcost}}
  \end{minipage}
  \begin{minipage}{0.45\linewidth}
  \subfigure[Average execution time]{
  \includegraphics[width=0.95\textwidth]{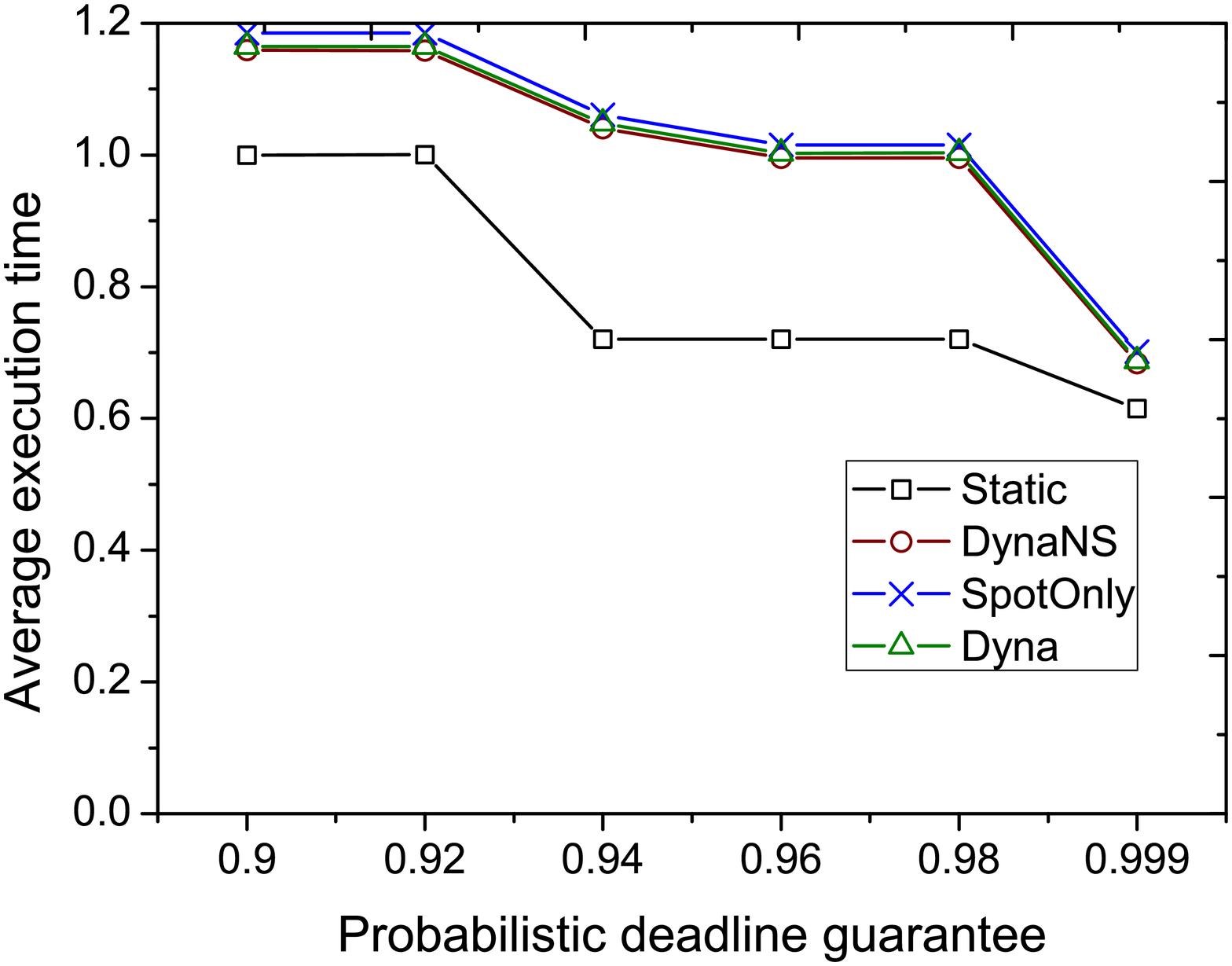}\label{fig:meetdltime}}
  \end{minipage}
  \caption{The average monetary cost and average execution time results of sensitivity studies on the probabilistic deadline guarantees.}\label{fig:meetdl}
\end{figure}

 \begin{table}
  \centering\begin{small}
  \caption{Obtained deadline hit rates by the compared algorithms with varying required deadline hit rate.}\label{tb:hitrate}
  \begin{tabular}{|m{2cm}|c|c|c|c|}
    \hline
    Required deadline hit rate&\multicolumn{4}{|c|}{Obtained deadline hit rate} \\ \cline{2-5}
    & Static & DynaNS & SpotOnly & Dyna \\\hline
    90\%&98.7\%&90.7\%&90.6\%&90.7\% \\\hline
    92\%&98.7\%&92.4\%&92.3\%&92.5\% \\\hline
    94\%&99.9\%&94.4\%&94.4\%&94.3\% \\\hline
    96\%&99.9\%&96.6\%&96.6\%&96.6\% \\\hline
    98\%&99.9\%&98.2\%&98.2\%&98.0\% \\\hline
    99.9\%&100\%&100\%&100\%&100\% \\\hline
\end{tabular}
\end{small}
\end{table}

{\bf Probabilistic deadline Guarantee.}
We evaluate the effectiveness of Dyna on satisfying probabilistic
deadline requirements when the requirement varies from 90\% to 99.9\%.
Figure~\ref{fig:meetdl} shows the average monetary cost per job and
average execution time results of the compared algorithms.
Dyna achieves the smallest monetary cost for different
probabilistic deadline guarantee settings. With a lower probabilistic
deadline requirement, the monetary cost saved by Dyna is higher.
Table~\ref{tb:hitrate} shows the obtained deadline hit rate by the compared algorithms with varying
required deadline hit rate. DynaNS, SpotOnly and Dyna can accurately
satisfy the probabilistic deadline guarantees while Static cannot.

\begin{figure}
  \centering
  \includegraphics[width= 0.3\textwidth]{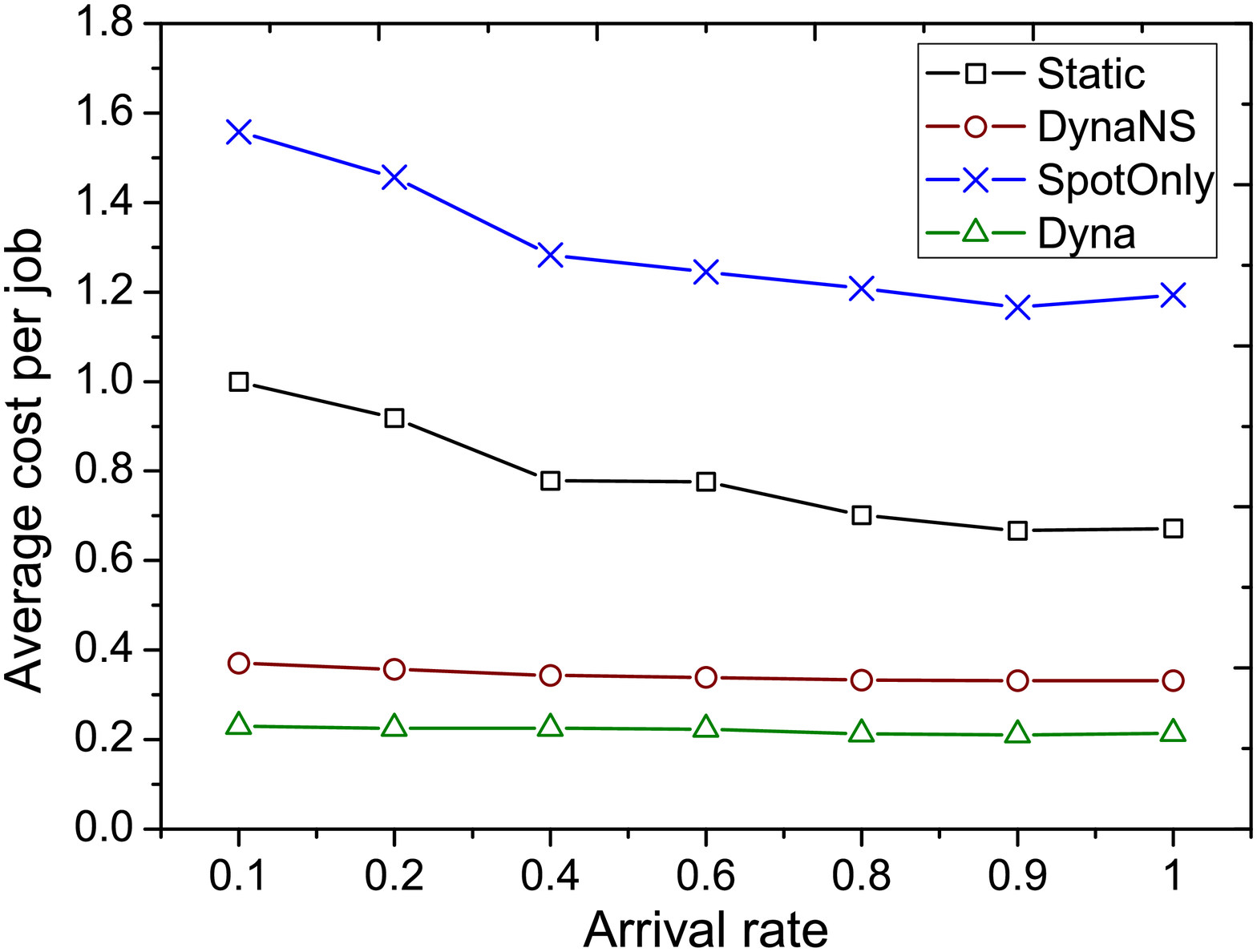}\label{fig:lambdacost}
  \centering\vspace{-0.1cm}
  \caption{The average monetary cost and average execution time results of sensitivity studies on the arrival rate of workflows.}\label{fig:lambda}
\end{figure}

{\bf Arrival rate.} We evaluate the effectiveness of Dyna when the arrival
rate $\lambda$ of workflows varies from 0.1, 0.2, 0.4, 0.6, 0.8, 0.9 to 1.0.
All results are normalized
to those when arrival rate is 0.1. Figure~\ref{fig:lambda}
shows the optimized average monetary cost per job.
Dyna obtains the smallest average monetary cost under all job arrival rates.
As the job arrival rate increases, the average cost per job is decreasing. This is
because there are more jobs sharing the resources rented by the WaaS provider from the cloud
and also sharing the hourly cost of instances charged by the IaaS cloud.

\begin{figure}
\centering
  \includegraphics[width= 0.3\textwidth]{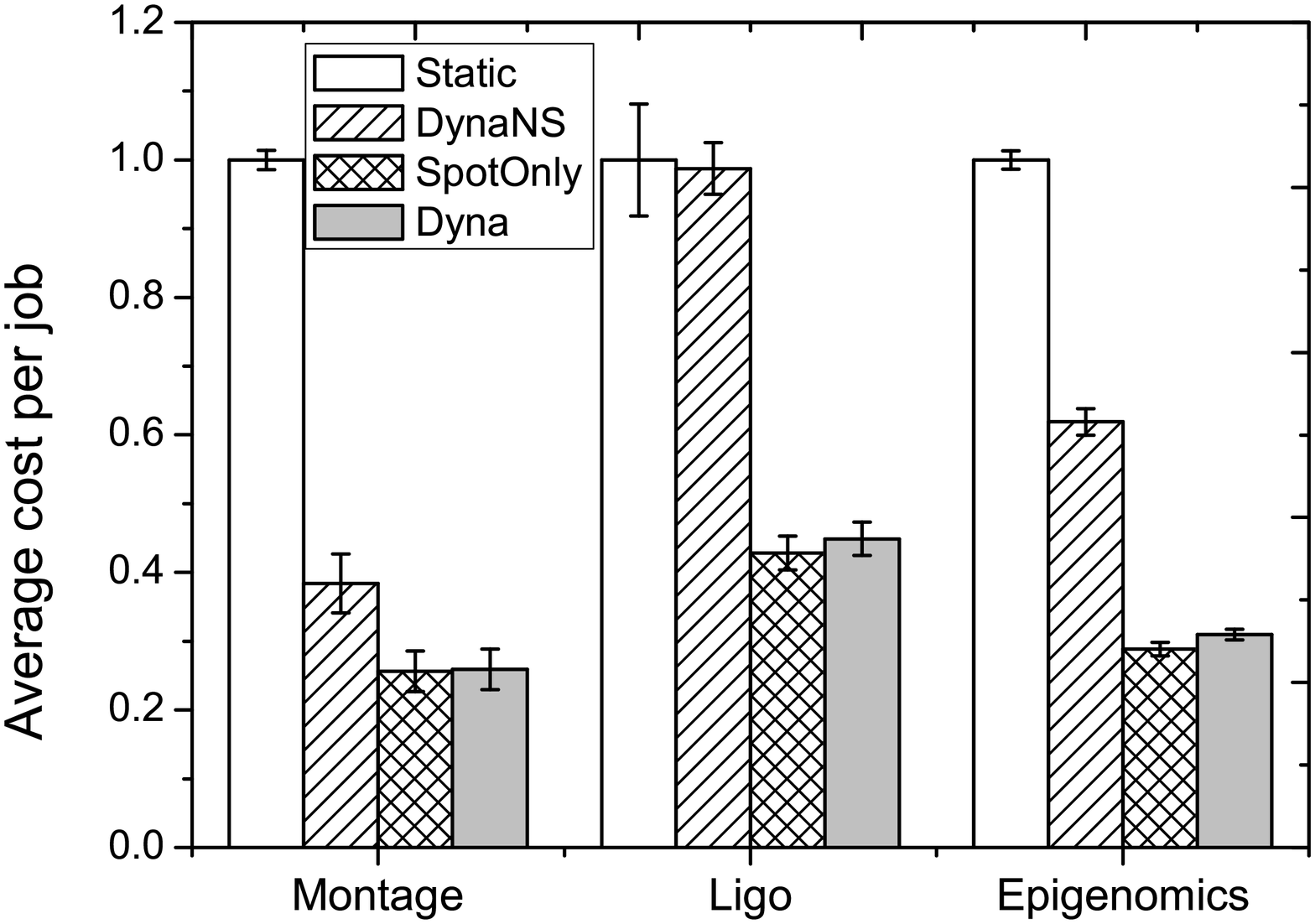}
\centering\vspace{-0.1cm}
  \caption{The simulation result of the average monetary cost obtained by the compared algorithms, using the spot price history of the Asia Pacific Region of Amazon EC2 in December, 2011.}
\label{fig:spotflat}       
\end{figure}

\begin{table}
  \centering\begin{small}
  \caption{Statistics on spot prices (\$/hour, December 2011, Asia Pacific Region) and on-demand prices of Amazon EC2. }\label{tb:spotflat}
  \begin{tabular}{|c|c|c|c|c|c|c|c|}
    \hline
    Instance type&Average&$stdev$&Min&Max&OnDemand\\\hline
    m1.small&0.041&0.003&0.038&0.05&0.06\\\hline
    m1.medium&0.0676&0.003&0.064&0.08&0.12\\\hline
    m1.large&0.160&0.005&0.152&0.172&0.24\\\hline
    m1.xlarge&0.320&0.009&0.304&0.336&0.48\\\hline
\end{tabular}
\end{small}
\end{table}

{\bf Spot price.} To study the sensitivity of Dyna and SpotOnly to the spot price variance, we
use simulations to study the compared algorithms on different spot price histories.
Particularly, we study the compared algorithms with the spot price history
of the Asia Pacific Region in
December 2011. As shown in Table~\ref{tb:spotflat},
the spot price during this period is very low and stable, in comparison with the period that we performed the experiments in August 2013.
Thus the spot instances are less likely to fail during the execution
(the failing probability $\mathit{ffp}$ is rather low).
We conjecture that more users are using the spot instances from Amazon EC2, which causes the larger fluctuations in August 2013 than December 2011.
Figure~\ref{fig:spotflat} shows the obtained monetary cost result.
SpotOnly and Dyna obtain similar monetary cost
results, which are much lower than Static and DynaNS.
This demonstrates Dyna is able to outperform SpotOnly on monetary
cost optimization for different spot price distributions.

%
%
%
%

\section{Conclusions}\label{sec:conclusion}
As the popularity of various scientific and data-intensive applications
in the cloud, hosting WaaS in IaaS clouds becomes emerging.
However, the IaaS cloud is a dynamic environment with performance and price dynamics,
which make the assumption of static task execution time and the QoS
definition of deterministic deadlines undesirable. In this paper, we
propose the notion of probabilistic performance guarantees as QoS
in dynamic cloud environments. We develop a probabilistic
framework named Dyna for scheduling scientific workflows with the
goal of minimizing the monetary cost while satisfying their
probabilistic deadline guarantees. The framework embraces a series
of static and dynamic optimizations. We further develop hybrid instance
configuration of spot and on-demand instances for price dynamics.
We deploy Dyna on Amazon EC2 and evaluate its effectiveness with real scientific workflows.
Our experimental results demonstrate that Dyna achieves much lower
monetary cost than the state-of-the-art approaches (by 73\%) while accurately meeting
users' probabilistic requirements.

\bibliographystyle{IEEEtran}
\begin{small}
\bibliography{spot}
\end{small}


\end{document}